\magnification=1000
\hoffset=1.2cm \voffset=0.2cm
\vbadness=10000

\font\SS=cmss10 scaled 1200

\font\ss=cmssq8 scaled 1200
\font\bf=cmbx10 scaled 1200
\font\bb=cmbx10 scaled 1728

\font\rm=cmr10 scaled 1440
\font\bbrm=cmr10 scaled\magstep2 

\font\it=cmti10 scaled 1200
\font\sl=cmsl10 scaled 1200
\font\sc=cmcsc10 scaled 1200
\font\tenrm=cmr10 scaled 1200
\font\ninerm=cmr9 scaled 1000
\font\eightrm=cmr8 scaled 1000
\font\sevenrm=cmr7 scaled 1000
\font\fiverm=cmr5 scaled 1000
\font\teni=cmmi10 scaled 1200
\font\seveni=cmmi7 scaled 1000
\font\fivei=cmmi5 scaled 1000
\font\sevensy=cmsy7 scaled 1000
\font\tensy=cmsy10 scaled 1200
\font\fivesy=cmsy5 scaled 1000

\font\tenbf=cmbx10 scaled 1200
\font\sevenbf=cmbx7 scaled 1200
\font\fivebf=cmbx5 scaled 1200
\font\tensl=cmsl10 scaled 1200
\font\tentt=cmtt10 scaled 1200
\font\tenit=cmti10 scaled 1200
\catcode`\@=11
\textfont0=\tenrm \scriptfont0=\sevenrm \scriptscriptfont0=\fiverm
\def\rm{\fam\z@\tenrm}
\textfont1=\teni \scriptfont1=\seveni \scriptscriptfont1=\fivei
\def\mit{\fam\@ne} \def\oldstyle{\fam\@ne\teni}
\textfont2=\tensy \scriptfont2=\sevensy \scriptscriptfont2=\fivesy
\def\cal{\fam\tw@}
\textfont3=\tenex \scriptfont3=\tenex \scriptscriptfont3=\tenex
\newfam\itfam \def\it{\fam\itfam\tenit} 
\textfont\itfam=\tenit
\newfam\slfam \def\sl{\fam\slfam\tensl} 
\textfont\slfam=\tensl
\newfam\bffam \def\bf{\fam\bffam\tenbf} 
\textfont\bffam=\tenbf \scriptfont\bffam=\sevenbf
\scriptscriptfont\bffam=\fivebf
\newfam\ttfam  
\textfont\ttfam=\tentt
\catcode`\@=12
\rm

\hfuzz=10pt \overfullrule=0pt
\vsize=8.9in
\hsize=14.9cm
\baselineskip=14pt

\def\singlespace{\baselineskip=14pt}
\parindent 20pt \parskip 1pt

\def\ctrline#1{\centerline{#1}}
\def\linebreak{\hfil\break}
\def\blankline{\par\vskip \baselineskip}
\def\twocol#1{\halign{##\quad\hfil &##\hfil\cr #1}}

 \mathcode`*="002A

\def\oforder{\sim}

\def\defeq{\equiv}

\def\neq{\not=}
\def\<={\leq}
\def\>={\geq}
\def\lsls{\ll}
\def\grgr{\gg}

\def\^{\char'017{}}
\def\%{\char'045{}}
\def\_{\vrule height 0.8pt depth 0pt width 1em}

\let\Mdollar=\$
\def\${\ifmmode\Mdollar\else$\Mdollar$\fi}

\def\spose\rlap
\def\caret#1{\widehat #1}

\def\cfalh{\par\vfil\eject \vskip -12pt \moveleft 0.5in\vbox{
     \twocol{{\bb Center for Astrophysics}\hbox to 1.5in{} &\cr
     {\ss 60 Garden Street} & {\ss Harvard College Observatory} \cr
     {\ss Cambridge, Massachusetts 02138} & {\ss Smithsonian
          Astrophysical Observatory}\cr}}\par\blankline}
\def\physicslh{\par\vfil\eject \vskip -12pt \vbox{
     \ctrline{\SS HARVARD UNIVERSITY}
     \vbox to 5 pt {}
     \hbox to \hsize{{\ss Department of 
Physics}\hfil {\ss Lyman Laboratory
          of Physics}}
     \hbox to \hsize{\hfil {\ss Cambridge, Massachusetts 02138}}}\par\blankline}

\def\date#1{\par\hbox to \hsize{\hfil #1\qquad}\par}

\def\ref{\par\noindent\hangindent 20pt}
\def\title#1\endtitle{\par\vfil\eject
     \par\vbox to 1.5in {}{\bf #1}\par\vskip 1.5in\nobreak}
\def\author#1\endauthor{\par\sc#1\par\blankline}

\def\sect#1\endsect{\par\vfil\eject{\bf #1}\par\vskip 12pt\nobreak}
\def\subsect#1\endsubsect{\vskip 14pt plus 50pt {\bf #1}\par
     \nobreak\blankline\nobreak}

\headline={\hss\tenrm\hss}
\singlespace
\nopagenumbers

\hbox{ }
\vskip 2.8cm
\ctrline{\bbrm Magnetohydrodynamic Density Waves in a Composite}
\vskip 0.4cm
\ctrline{\bbrm Disk System of Interstellar Medium and Cosmic-Ray Gas}
\vskip 12pt
\vskip 1.0cm
\ctrline{{\rm Y}{\ninerm U}-{\rm Q}{\ninerm ING} 
{\rm L}{\ninerm OU}$^{1,2,3}$ {\ninerm AND}
{\rm Z}{\ninerm UHUI} {\rm F}{\ninerm AN}$^{4}$ }
\vskip 1.0cm
{\sl
\hskip 2.8cm $^1$National Astronomical Observatories

\hskip 3.5cm Chinese Academy of Sciences

\hskip 3.5cm A20 Datun Road, ChaoYang District, 

\hskip 3.5cm Beijing, 100012, China }
\vskip 0.5cm

{\sl
\hskip 2.8cm $^2$Physics Department, Tsinghua Astrophysics Center,

\hskip 3.5cm Tsinghua University, Beijing, 100084, China }
\vskip 0.5cm

{\sl
\hskip 2.8cm $^3$Department of Astronomy and Astrophysics 

\hskip 3.5cm The University of Chicago

\hskip 3.5cm Chicago, Illinois 60637, USA

\hskip 3.5cm lou@oddjob.uchicago.edu }
\vskip 0.5cm

{\sl
\hskip 2.8cm $^4$Beijing Astrophysical Center and

\hskip 3.5cm Department of Astronomy, CAS-PKU,

\hskip 3.5cm Beijing, 100871, China } 

\vskip 1.0cm


\par\vfill\eject

\magnification=1000
\hoffset=1.2cm \voffset=0.2cm
\vbadness=10000

\font\SS=cmss10 scaled 1200

\font\ss=cmssq8 scaled 1200
\font\bf=cmbx10 scaled 1200
\font\bb=cmbx10 scaled 1728

\font\rm=cmr10 scaled 1440
\font\bbrm=cmr10 scaled\magstep2 

\font\it=cmti10 scaled 1200
\font\sl=cmsl10 scaled 1200
\font\sc=cmcsc10 scaled 1200
\font\tenrm=cmr10 scaled 1200
\font\ninerm=cmr9 scaled 1200
\font\eightrm=cmr8 scaled 1200
\font\sevenrm=cmr7 scaled 1000
\font\fiverm=cmr5 scaled 1000
\font\teni=cmmi10 scaled 1200
\font\seveni=cmmi7 scaled 1000
\font\fivei=cmmi5 scaled 1000
\font\sevensy=cmsy7 scaled 1000
\font\tensy=cmsy10 scaled 1200
\font\fivesy=cmsy5 scaled 1000

\font\tenbf=cmbx10 scaled 1200
\font\sevenbf=cmbx7 scaled 1200
\font\fivebf=cmbx5 scaled 1200
\font\tensl=cmsl10 scaled 1200
\font\tentt=cmtt10 scaled 1200
\font\tenit=cmti10 scaled 1200
\catcode`\@=11
\textfont0=\tenrm \scriptfont0=\sevenrm \scriptscriptfont0=\fiverm
\def\rm{\fam\z@\tenrm}
\textfont1=\teni \scriptfont1=\seveni \scriptscriptfont1=\fivei
\def\mit{\fam\@ne} \def\oldstyle{\fam\@ne\teni}
\textfont2=\tensy \scriptfont2=\sevensy \scriptscriptfont2=\fivesy
\def\cal{\fam\tw@}
\textfont3=\tenex \scriptfont3=\tenex \scriptscriptfont3=\tenex
\newfam\itfam \def\it{\fam\itfam\tenit} 
\textfont\itfam=\tenit
\newfam\slfam \def\sl{\fam\slfam\tensl} 
\textfont\slfam=\tensl
\newfam\bffam \def\bf{\fam\bffam\tenbf} 
\textfont\bffam=\tenbf \scriptfont\bffam=\sevenbf
\scriptscriptfont\bffam=\fivebf
\newfam\ttfam  
\textfont\ttfam=\tentt
\catcode`\@=12
\rm

\hfuzz=10pt \overfullrule=0pt
\vsize=8.9in
\hsize=14.9cm
\baselineskip=12pt

\def\singlespace{\baselineskip=13pt}
\parindent 20pt \parskip 6pt

\def\ctrline#1{\centerline{#1}}
\def\linebreak{\hfil\break}
\def\blankline{\par\vskip \baselineskip}
\def\twocol#1{\halign{##\quad\hfil &##\hfil\cr #1}}

 \mathcode`*="002A

\def\oforder{\sim}

\def\defeq{\equiv}

\def\neq{\not=}
\def\<={\leq}
\def\>={\geq}
\def\lsls{\ll}
\def\grgr{\gg}

\def\^{\char'017{}}
\def\%{\char'045{}}
\def\_{\vrule height 0.8pt depth 0pt width 1em}

\let\Mdollar=\$
\def\${\ifmmode\Mdollar\else$\Mdollar$\fi}

\def\spose\rlap
\def\caret#1{\widehat #1}

\def\cfalh{\par\vfil\eject \vskip -12pt \moveleft 0.5in\vbox{
     \twocol{{\bb Center for Astrophysics}\hbox to 1.5in{} &\cr
     {\ss 60 Garden Street} & {\ss Harvard College Observatory} \cr
     {\ss Cambridge, Massachusetts 02138} & {\ss Smithsonian
          Astrophysical Observatory}\cr}}\par\blankline}
\def\physicslh{\par\vfil\eject \vskip -12pt \vbox{
     \ctrline{\SS HARVARD UNIVERSITY}
     \vbox to 5 pt {}
     \hbox to \hsize{{\ss Department of 
Physics}\hfil {\ss Lyman Laboratory
          of Physics}}
     \hbox to \hsize{\hfil {\ss Cambridge, Massachusetts 02138}}}\par\blankline}

\def\date#1{\par\hbox to \hsize{\hfil #1\qquad}\par}

\def\ref{\par\noindent\hangindent 20pt}
\def\title#1\endtitle{\par\vfil\eject
     \par\vbox to 1.5in {}{\bf #1}\par\vskip 1.5in\nobreak}
\def\author#1\endauthor{\par\sc#1\par\blankline}

\def\sect#1\endsect{\par\vfil\eject{\bf #1}\par\vskip 12pt\nobreak}
\def\subsect#1\endsubsect{\vskip 14pt plus 50pt {\bf #1}\par
     \nobreak\blankline\nobreak}

\pageno=2
\headline={\hss\tenrm\folio\hss}
\def\gsim{\;\lower4pt\hbox{${\buildrel\displaystyle >\over\sim}$}\;}
\def\lsim{\;\lower4pt\hbox{${\buildrel\displaystyle <\over\sim}$}\;}
\singlespace
\nopagenumbers

\hbox{ }
\ctrline{ ABSTRACT }
\vskip 0.4cm
Multi-wavelength observations from radio to soft X-ray bands 
of large-scale galactic spiral structures offer synthesized and
comprehensive views of nearby disk galaxies. In the presence of 
a massive dark-matter halo, the density-wave dynamics on galactic 
scales involves the stellar disk, the gas disk of interstellar 
medium (ISM), the magnetic field, and the cosmic-ray gas (CRG). In 
this paper, we explore the dynamic and electromagnetic interplay 
between the magnetized ISM disk and CRG disk so that structural 
and diagnostic features of optical, infrared, and synchrotron 
radio-continuum emissions from a spiral galaxy can be physically 
understood. On timescales of galactic density waves, cosmic rays 
collectively may be treated as a relativistically hot tenuous gas 
fluid that is tied to the large-scale mean magnetic field in 
transverse bulk motions but moves otherwise differently along the 
magnetic field relative to the ISM. For both fast and slow 
magnetohydrodynamic (MHD) density waves in a composite disk 
system of magnetized ISM and CRG, the minute CRG mass density 
enhancement is phase shifted relative to the enhancement of parallel 
magnetic field. Owing to an extremely small number of cosmic rays, 
the large-scale magnetic field enhancement dominates in synchrotron 
radio-continuum emissions (as if the CRG is almost unperturbed) for 
spiral structural manifestations. In addition to the fast and 
slow MHD density waves, there also exists a suprathermal MHD wave 
mode by which CRG adjusts itself with an effective suprathermal 
sound speed close to the speed of light $c$.
\vskip 0.3cm
\noindent
Key words: galaxies: cosmic rays --- density waves --- radio polarization  
--- gravitation --- magnetic fields --- interstellar medium 
\vfill\eject

\hbox{ }
\vskip 0.4cm
\ctrline{1. INTRODUCTION}
\vskip 0.4cm
The entire system of a spiral galaxy consists of a massive dark-matter 
halo, an older stellar halo, a luminous stellar disk with an central 
ellipsoidal bulge, a thin disk of interstellar medium (ISM) consisting 
of a thermal gas plus dusts, relativistic cosmic rays, and a large-scale
magnetic field (e.g., Woltjer 1965). The rotation curve and speed 
of a disk galaxy are dominantly controlled by the total mass (dark 
matter included) distribution in the system via gravity (e.g., Kormendy 
\& Norman 1979; Kent 1986, 1987, 1988). The spiral structure in the disk 
seen in red light from the relatively old stellar population is usually 
broad and smooth (e.g. Elmegreen 1981), while the spiral arm structure 
in the disk seen in blue light from the young stellar population is 
typically brilliant, narrow, and sharp. Often, there exist dark narrow 
dust lanes lying along inner edges of the 
luminous spiral arms outlined by young O, B stars and H{\eightrm II} 
complexes. The large-scale spiral structure seen in total synchrotron 
radio-continuum emissions well tracks the optical spiral structure 
(e.g., the ``Whirlpool galaxy" M51 [NGC 5194]; Mathewson, van der 
Kruit, \& Brouw 1972; Neininger 1992; Berkhuijsen et al. 1997). 
Polarized radio-continuum emission arms also follow optical 
spiral arms as seen in several nearby spiral galaxies (e.g. M51, 
M31, NGC 2997). However, the spiral structure seen in polarized 
synchrotron radio-continuum emissions can, in the case of the nearby 
spiral galaxy NGC 6946 (Beck \& Hoernes 1996; Fan \& Lou 1996, 1999; 
Lou \& Fan 1998a, 2002; Ferguson et al. 1998; Frick et al. 2000; 
Lou 2002), appear interlaced with the optical spiral 
structure.\footnote{$^1$}{Two other late-type gas-rich spiral galaxies 
nearby, IC342 (Krause 1993; Crosthwatte et al. 2000) and M83 or NGC 5236 
(Sukumar \& Allen 1989; Neininger, Beck, Sukumar, \& Allen 1993), also 
bear somewhat similar interlaced arm features of NGC 6946.} 
Synchrotron radio-continuum emissions are produced by relativistic 
cosmic-ray electrons gyrating around magnetic fields. The 
large-scale interrelations between optical and radio-continuum spiral 
structures seen in spiral galaxies hint at an intricate magnetohydrodynamic 
(MHD) coupling among various seemingly unrelated processes on distinctly 
different scales (Lou \& Fan 1997, 1998b, 2000a, b; Lou 2002).

In addition to giving rise to a more or less flat disk rotation curve, 
the massive dark-matter halo also helps prevent rapid development of 
bar-type instabilities that have been shown to exist through numerical 
and theoretical studies (Miller, Prendergast, \& Quirk 1970; Hohl 1971; 
Ostriker \& Peebles 1973; Bardeen 1975; Shu et al. 2000). 
By joint effects of differential rotation, epicyclic oscillations, 
and self-gravity, such a disk system is still vulnerable to various 
instabilities (Safronov 1960; Toomre 1964; Goldreich \& Lynden-Bell 
1965; Julian \& Toomre 1966; Jog \& Solomon 1984a, b; Binney \& 
Tremaine 1987; Bertin et al. 1989a, b; Bertin \& Lin 1996; Montenegro 
et al. 1999; Lou, Yuan, \& Fan 2001). In the case of a stellar disk, 
such instabilities may ultimately lead to an effective increase of 
stellar velocity dispersion. In the case of a rotating gas disk, such 
instabilities may trigger cloud and star formations on global scales 
(Quirk 1972; Jog \& Solomon 1984a, b; Kennicutt 1989; Kennicutt et al. 
1994; Wang \& Silk 1994; Elmegreen 1995; Jog 1992, 1996; Silk 1997; 
Lou \& Fan 1998b). As clouds and stars directly form in the gas disk of 
interstellar medium (ISM) and the magnetic and thermal energy densities 
in the ISM are roughly comparable, magnetic field should play an important 
role in affecting the global star formation rate. In particular, magnetic 
field introduces {\it additional} MHD instabilities (Lou \& Fan 1997, 
1998a, 2000a, b; Lou et al. 2001a) that are independent of Toomre-type 
ring instabilities for axisymmetric disturbances and that can lead to 
nonaxisymmetric gravitational collapses (Lou 1996a; Lou et al. 2001a).

Over the past decade, the advent of Infrared Astronomical 
Satellite (IRAS) and Infrared Space Observatory (ISO) has opened 
up the infrared window for systematic, extensive, 
and unprecedented galactic observations. One important discovery of 
IRAS was the remarkably tight {\it global correlation} between 
integrated far-infrared and radio-continuum emissions from spiral 
galaxies (Dickey \& Salpeter 1984; Helou, Soifer \& Rowan-Robinson 
1985; de Jong et al. 1985; Wunderlich, Klein \& Wielebinski 1987; 
Bicay \& Helou 1990; Helou 1991). This empirical fact implies a 
close physical connection between two apparently unrelated physical 
mechanisms, namely, thermal emissions from UV photon heated dusts 
in the ISM and synchrotron radio-continuum emissions from relativistic 
cosmic-ray electrons gyrating around galactic magnetic field. 
It was proposed that massive star formation in the ISM is likely the 
key process that somehow links these disparate radiative phenomena 
(Harwit \& Pacini 1975; Helou et al. 1985; Condon, Anderson, \& 
Helou 1991). Recent high-resolution ISO observations of the nearby 
spiral galaxes NGC 6946 (e.g. Tuffs et al. 1996; Lu et al. 1996; Helou 
et al. 1996; Malhotra et al. 1996) and M31 (see extensive references in 
Berkhuijsen, Beck \& Walterbos 2000) further revealed detailed spatial 
correlations and correspondences between large-scale infrared and radio 
structures as already partially evidenced in an earlier IRAS survey of 
spiral galaxies (Bicay \& Helou 1990). We emphasize the importance of 
multi-wavelength observations of spiral galaxies (Lou, Walsh, Han \& Fan 
2002), because we believe that large-scale spiral MHD density waves are 
the underlying dynamic process that links and organizes the various ISM 
components through magnetic field and that gives rise to large-scale 
interrelations of spiral structures revealed in various wavebands 
(Lou \& Fan 2000a, b).
 
The familiar hydrodynamic density wave theory (Lin \& Shu 1964, 1966) 
dealt primarily with large-scale perturbations in a thin stellar disk, 
using either a formalism of stellar distribution function (e.g., Toomre 
1964; Julian \& Toomre 1966) or a fluid approach (e.g., Lin 1967b, 1987;
Toomre 1977; Binney \& Tremaine 1987; Bertin \& Lin 1996). The inclusion 
of a less massive ISM disk was, in the past, regarded as a passive effect. 
Nevertheless, there has been growing evidence for the important dynamic 
as well as diagnostic roles of 
the ISM (e.g., Jog \& Solomon 1984a, b). While smaller than the kinetic 
energy density of the galactic disk rotation, the energy densities of 
thermal ISM, magnetic field, and cosmic rays are comparable (e.g.,
Lin 1967a). The increase of dynamical freedoms in the magnetized ISM 
may lead to different large-scale structural interrelations in 
{\it multi-wavelength} observations and the presence of magnetic field 
would significantly influence the global star formation rate in a 
nontrivial manner. Therefore, in order to understand large-scale 
infrared and synchrotron radio-continuum structures of spiral galaxies 
as well as their interrelations with the optical spiral structures, it 
becomes necessary to take into account the effects of thermal ISM, 
magnetic field, and cosmic rays within the overall MHD density wave 
scenario. While the gravitational effect dominates the dynamics of the 
massive stellar disk and of the mutual interaction between the stellar 
and ISM disks (Lou \& Fan 1997, 1998b, 2000a, b), physical consequences 
of the dynamical interaction among the sub-systems remain to be 
specifically and thoroughly examined.

While the ultimate goal of building up such an edifice remains 
challenging, it is nonetheless physically informative to extract 
necessary ingredients separately from such a comprehensive scenario 
in order to work out simpler yet non-trivial physical partial problems 
in concrete terms. This research strategy requires a systematic 
investigation on a series of partial problems involving various 
simplifications. Meanwhile, one must clearly bear in mind the utility 
as well as limitations of the theoretical results thus derived. We 
hope to gain physical insights for the overall problem such that 
various aspects can be modeled quantitatively to confront 
multi-wavelength observations of spiral galaxies (Lou et al. 2002). 

We have shown in earlier publications the possible existence of fast 
and slow MHD density 
waves\footnote{$^2$}{Note that incompressible Alfv\'enic fluctuations 
perpendicular to a magnetized rotating gas disk may exist (sort of
bending MHD waves), but they do not directly couple to gravitational 
potential perturbations in the absence of mass density fluctuations 
(see the formulation contained in \S 2 and in Appendix C).} 
in a differentially rotating thin gas disk embedded with an 
azimuthal magnetic field 
(Fan \& Lou 1996; Lou \& Fan 1998a). In particular, perturbation 
enhancements of thermal gas density and parallel magnetic field are 
roughly in phase for fast MHD density waves but are significantly 
phase shifted for slow MHD density waves with a phase difference 
$\gsim\pi/2$. The features of fast MHD density waves may explain why 
optical and synchrotron radio-continuum structures nearly coincide in 
several nearby spiral galaxies (e.g. M51, M31, NGC 2997 etc.),
whereas the features of slow MHD density waves may explain why magnetic 
spiral arms lie in between the optical spiral arms in the galaxy 
NGC 6946 and why such an interlaced spiral structure persists
over a large spatial scale (Beck \& Hoernes 1996; Fan \& Lou 1996; 
Ferguson et al. 1998; Frick et al. 2000; Lou \& Fan 1998a, 2002).

An important new realm of applications for spiral MHD density 
wave theory has been developed recently for {\it circumnuclear 
spiral arms and starburst ``rings"} on kiloparsec scales in the 
central regions around nuclei of disk galaxies such as NGC 1097, 
NGC 6951, NGC 2997, and NGC 2207 etc. (Lou, Yuan, Fan, \& Leon 2001). 
More specifically, when coupled with proper wave damping mechanisms
in a circumnuclear disk, spiral MHD density waves play the key role of 
persistently removing angular momentum from the central magnetized 
gas disk in the circumnuclear region, and the incessant accumulation 
of gas materials {\it and} magnetic flux in a roughly circular zone 
is eventually vulnerable to gravitational instabilities (e.g., 
Elmegreen 1994; 
Lou 1996a) and thus leads to the appearance of a starburst ``ring" 
somewhere inside the modified inner Lindblad resonance (Lou et al. 
2001b). At the current stage of observations, wavelet analyses and 
reconstructions on high-resolution optical images from {\it Hubble 
Space Telescope (HST)} have revealed circumnuclear spiral structures 
with or without nuclear bars in several dozens of spiral galaxies, 
including NGC 2207 (Yuan, private communications 2001). The starburst 
galaxy M83 (NGC 5236) has been revealed to possess double circumnuclear 
ring and minibar on scales of several hundred parsecs (Elmegreen et al. 
1998). In comparison, preliminary information of circumnuclear magnetic 
fields are only known recently for NGC 1097 (Beck et al. 1999) and barely 
for NGC 2997 (Han et al. 1999; Lou et al. 1999). Here again, it is the 
relativistic cosmic-ray electrons abundant in the circumnuclear environs 
that reveal valuable clues of trailing swirl patterns of circumnuclear 
magnetic fields.

The central theme of this paper is to investigate the coupling of spiral 
MHD density waves in the magnetized thermal ISM disk and the cosmic-ray 
gas (CRG). We are mainly motivated by the fact that the energy densities of 
thermal gas, magnetic field, and CRG are comparable (all on the order of 
$\sim 10^{-12}\hbox{ erg cm}^{-3}$), the fact that galactic synchrotron 
radio-continuum emissions result from interactions of relativistic 
cosmic-ray electrons with the magnetic field, and the fact that magnetic 
field appears to be the underlying cause for the large-scale structural 
correlations as revealed by multi-wavelength observations of nearby
spiral galaxies. We do not include the effect of the massive stellar 
disk in the formulation merely for the sake of simplicity; yet we have 
enough confidence, based on our earlier analyses (Lou \& Fan 1997, 1998b, 
2000a, b), that the gross properties of fast and slow MHD density waves 
should remain in the magnetized thermal gas disk of ISM even when the 
more massive stellar disk is included (see Lou et al. 2001a; Lou \&
Shen 2003).

Our basic theoretical formalism of the interaction between the 
magnetized thermal gas and CRG closely follows that pioneered by 
Parker (1965, 1967, 1969) nearly four decades ago. The key starting 
point is to treat galactic cosmic rays collectively as a tenuous 
relativistically hot gas fluid for dynamic phenomena of large spatial 
scales (i.e. $\gg$ gyroradii of cosmic rays) and low frequencies.
In this paper, we therefore ignore resistive effects such that magnetic
field is frozen into both ISM and CRG. The physical meaning of frozen-in 
condition is that bulk materials cannot go across magnetic field lines 
but are allowed to move along magnetic field lines. As ISM and CRG are 
two qualitatively different types of conducting fluids, their bulk 
motions along magnetic field lines are different following the proper 
dynamics. In other words, both thermal ISM and CRG are tied to the 
galactic magnetic field in transverse bulk motions but are allowed to 
move relative to each other along magnetic field lines. By this analysis, 
it is then possible to assess the relative importance of the enhancements 
of parallel magnetic field and CRG mass density in synchrotron 
radio-continuum structures of spiral galaxies.

The plan of the paper is as follows. Physical considerations
and perturbation formulation of the problem are developed in \S 2. 
Dispersion relations of fast and slow MHD density waves and of the 
suprathermal MHD mode are derived in \S 3 using the tight-winding 
approximation. Galactic applications are described in \S 4. Notes 
and discussions are contained in \S 5. Appendix A contains some 
mathematical details of the key perturbation equation (3.12) for 
the convenience of reference and discussion. In Appendix B, we 
derive and summarize specific phase relationships of perturbation 
variables for the problem studied by Parker (1965) yet with the 
displacement current effect included. In Appendix C, we provide 
a first-principle formulation as the very basis of our MHD 
perturbation analysis.

\hbox{ }
\vskip 0.4cm
\ctrline{2. FORMULATION OF THE PROBLEM }
\vskip 0.4cm

For the mathematical description of the problem under consideration
(see Appendix C for more specifics), we adopt the cylindrical 
coordinate system $(r,\theta,z)$ with the $z$- and disk rotation 
axes being coincident. For the stationary rotational equilibrium
of the background, 
we presume the bulk azimuthal speed $U_{\theta}$ of the CRG to be 
the same as the bulk azimuthal speed $V_{\theta}$ of the thermal 
gas of ISM.\footnote{$^3$}{In general, $V_{\theta}$ and $U_{\theta}$
may be different. In galactic contexts, this difference should be
small, because the total gravity (that of dark matter halo included)
dominates over pressure and Lorentz forces. Furthermore, 
Kelvin-Helmholtz instabilities induced by small differences in 
$V_{\theta}$ and $U_{\theta}$ may sustain a certain level of 
microturbulence that tends to smooth out such velocity shear. }
The background magnetic field $B_{\theta}$ is azimuthal 
to avoid the magnetic field winding dilemma due to the disk 
differential rotation (Lynden-Bell 1966; Roberts \& Yuan 1970). 
The profile of the background magneto-rotational equilibrium is 
characterized by $U_{\theta}=V_{\theta}=\Omega r$ and 
$B_{\theta}=F_B/r$, where $\Omega (r)$ is the angular rotation rate 
and $F_B$ is a constant. The stationary radial momentum equation 
for the background becomes 
$$
-{(\mu_{\circ}+\epsilon_{\circ})V_{\theta}^2\over r}
=-{\partial (p_{\circ}+P_{\circ})\over\partial r}
-\int dz{B_{\theta}\over 4\pi r}{\partial (rB_{\theta})\over\partial r}
+(\mu_{\circ}+\epsilon_{\circ}){\partial\phi_{T}\over\partial r}\ 
\eqno(2.1)
$$
(Lou 1996b; see Appendix C), where $\mu_{\circ}$ is the surface mass density
of the thermal gas, $p_{\circ}$ is the two-dimensional thermal gas 
pressure, $\phi_T$ is the {\it total} negative gravitational potential, 
$\epsilon_{\circ}\defeq\Sigma_{\circ}+\Gamma P_{\circ}/[(\Gamma-1)c^2]$, 
$\Sigma_{\circ}$ is the vertically integrated mass density of the CRG, 
$P_{\circ}$ is the two-dimensional effective CRG pressure taken to be 
isotropic (Parker 1969), $c$ is the speed of light, and $\Gamma$ is the 
polytropic index of the CRG (e.g., $\Gamma=4/3$ in a relativistically 
hot gas). The axisymmetric Poisson equation is 
$$
{1\over r}{\partial\over\partial r}
\bigg(r{\partial\phi_{T}\over\partial r}\bigg)
+{\partial^2\phi_{T}\over\partial z^2}=-4\pi\rho_{T}\ ,\eqno(2.2)
$$
with $\phi_{T}$ felt at the magnetized gas disk being determined 
by the distribution of the {\it total} mass density $\rho_{T}$ 
in the entire system. Within the solar 
system, the CRG was observed to be remarkably isotropic with 
an upper bound on anisotropy of $\sim 10^{-3}$ (Greisen 1960). In general, 
small-scale plasma instabilities are expected to reduce the anisotropy of 
the CRG to be $\lsim 1\%$ (Lerche \& Parker 1966; Lerche 1967). 
Thermal gas pressure, effective CRG pressure, and magnetic energy 
density $B_{\theta}^2/(8\pi)$ are roughly comparable and are
on the order of $\oforder 10^{-12}
\hbox{ erg cm}^{-3}$. 

The thickness of the composite disk system is expediently taken to be 
infinitely thin here. In reality, the vertical quasi-static balance involves 
the confinement of the thermal ISM, CRG, and magnetic field under 
the action of gravity towards the galactic disk plane at $z=0$ from both
sides. In the background rotational equilibirum, the radial Lorentz force 
in equation (2.1) vanishes (i.e., 
force-free) as $rB_{\theta}$ is taken to be a constant $F_B$. By this choice 
of $B_{\theta}$ that scales as $r^{-1}$, one needs to invoke some processes 
in the vicinity of the galactic centre to avoid the outright singularity of 
$B_{\theta}$ there. As the massive dark-matter halo contributes to 
$\rho_{T}$ and thus to $\phi_{T}$, the disk rotation curve for the velocity 
$V_{\theta S}(r)$ of the {\it stellar disk} is actually used to infer 
$\phi_{T}$ from observations. Since the thermal energy density of the ISM 
is much smaller than the kinetic energy density of the galactic rotation 
and the magnetic field is taken to be force-free, the difference between 
the gas rotation velocity $V_{\theta}(r)$ and the stellar rotation 
velocity $V_{\theta S}(r)$ should be very small. In fact, one usually
prescribes a $V_{\theta}(r)$ a priori based on observational input to 
construct an approximate background profile.

Given a background profile for the thin rotating gas disk of magnetized 
ISM together with the CRG, it is fairly straightforward to write down MHD 
equations for large-scale coplanar perturbations, that is, we limit the 
consideration to two-dimensional propagations of MHD wave perturbations 
tangential to the disk plane at $z=0$. By this requirement, 
incompressible Alfv\'enic fluctuations involving velocity and magnetic 
field perturbations perpendicular to the disk plane are excluded 
(Lou \& Fan 1998a). Likewise, Parker instability (Parker 1966) and
magnetorotational instabilities (MRI; Balbus \& Hawley 1998; Kim \&
Ostriker 2000) are also excluded. In the context of spiral MHD density 
waves, we are mainly interested in {\it compressible fluctuations}, for 
which the relevant MHD perturbation equations are given below explicitly. 
The radial component of the magnetic induction equation for the 
radial magnetic field perturbation $b_r$ is
$$
{\partial b_r\over\partial t}=-{1\over r}{\partial (V_{\theta}b_r)
\over\partial\theta}+{1\over r}{\partial (B_{\theta}v_r)\over
\partial\theta}\ ,\eqno(2.3)
$$
where $v_r$ is the bulk radial thermal gas velocity perturbation, 
and the azimuthal magnetic induction equation for 
the azimuthal magnetic field perturbation $b_{\theta}$ is
$$
{\partial b_{\theta}\over\partial t}={\partial (V_{\theta}b_r)
\over\partial r}-{\partial (B_{\theta}v_r)\over\partial r}\ .\eqno(2.4)
$$
The divergence-free condition of the magnetic field 
perturbation $\vec b\defeq (b_r,b_{\theta},0)$ is
$$
{1\over r}{\partial (rb_r)\over\partial r}
+{1\over r}{\partial b_{\theta}\over\partial\theta}=0\ .\eqno(2.5)
$$
While the bulk radial velocity $u_r$ of the CRG is the same as 
$v_r$ as restricted by the frozen-in magnetic field on large scales, 
the bulk azimuthal flow speed $u_{\theta}$ of the CRG and the bulk 
azimuthal flow speed $v_{\theta}$ of the thermal gas can however 
be different along magnetic field lines.

The radial component of the perturbed momentum equation is
$$\eqalign{
{\partial v_r\over\partial t}
&-{2\mu_{\circ}V_{\theta}v_{\theta}\over r(\mu_{\circ}+\epsilon_{\circ})}
-{2\epsilon_{\circ}V_{\theta}u_{\theta}\over r(\mu_{\circ}+\epsilon_{\circ})}
+{V_{\theta}\over r}{\partial v_r\over\partial\theta}
-{(\mu+\epsilon)\over (\mu_{\circ}+\epsilon_{\circ})^2}
{d(p_{\circ}+P_{\circ})\over dr}\cr &
=-{1\over (\mu_{\circ}+\epsilon_{\circ})}
{\partial (p+P)\over\partial r}
-\int {dzB_{\theta}\over 4\pi r(\mu_{\circ}+\epsilon_{\circ})}
\bigg[{\partial (rb_{\theta})
\over\partial r}-{\partial b_r\over\partial\theta}\bigg]+
{\partial\phi\over\partial r}\ \cr}\eqno(2.6)
$$
(Lou 1996b; see Appendix C), where $\mu$, $\epsilon$, $p$, 
and $P$ are perturbations in $\mu_{\circ}$, $\epsilon_{\circ}$, 
$p_{\circ}$, and $P_{\circ}$, respectively, and $\phi$ is the 
negative self-gravity potential perturbation associated with the 
thermal ISM disk. The azimuthal momentum equation for the thermal 
ISM is
$$
{\partial v_{\theta}\over\partial t}
+{V_{\theta}\over r}{\partial v_{\theta}\over\partial\theta}
+{1\over r}{d (rV_{\theta})\over dr}v_r=-{1\over\mu_{\circ}r}
{\partial p\over\partial\theta}+{1\over r}{\partial\phi
\over\partial\theta}\ ,\eqno(2.7)
$$
and in parallel, the perturbed azimuthal 
momentum equation for the CRG is
$$
{\partial u_{\theta}\over\partial t}
+{V_{\theta}\over r}{\partial u_{\theta}\over\partial\theta}
+{1\over r}{d (rV_{\theta})\over dr}u_r=-{1\over\epsilon_{\circ}r}
{\partial P\over\partial\theta}+{1\over r}{\partial\phi
\over\partial\theta}\ .\eqno(2.8)
$$
The mass conservation for the surface mass 
density perturbation $\mu$ of the thermal ISM is
$$
{\partial\mu\over\partial t}+{1\over r}{\partial (\mu_{\circ}rv_r)
\over\partial r}+{\mu_{\circ}\over r}
{\partial v_{\theta}\over\partial\theta}+{V_{\theta}\over r}
{\partial\mu\over\partial\theta}=0\ .\eqno(2.9)
$$
The mass conservation for the mass 
density perturbation $\Sigma$ of the CRG is
$$
{\partial\Sigma\over\partial t}+{1\over r}
{\partial (\Sigma_{\circ}ru_r)
\over\partial r}+{\Sigma_{\circ}\over r}
{\partial u_{\theta}\over\partial\theta}+{V_{\theta}\over r}
{\partial\Sigma\over\partial\theta}=0\ .\eqno(2.10)
$$
The three-dimensional Poisson equation for the perturbed 
negative self-gravity potential $\phi$ induced by $\mu$ 
becomes 
$$
{1\over r}{\partial\over\partial r}
\bigg(r{\partial\phi\over\partial r}\bigg)
+{1\over r^2}{\partial^2\phi\over\partial\theta^2}
+{\partial^2\phi\over\partial z^2}
=-4\pi G\mu(r,\theta)\delta(z)\ ,\eqno(2.11)
$$
where the tiny contribution from the CRG to the negative 
self-gravity potential perturbation has been dropped, $\delta(z)$ 
is the Dirac delta function with argument $z$, and $\mu_{\circ}$ is 
related to the mass density of the thermal ISM $\rho_{\circ}$ 
by $\rho_{\circ}(r,z)\equiv\mu_{\circ}(r)\delta(z)$. For the 
magnetic field perturbation $\vec b$, one of equations (2.3) 
and (2.4) may be spared when equation (2.5) is used instead. The 
perturbed polytropic relation for the two-dimensional ISM pressure 
perturbation $p$ and the surface mass density perturbation $\mu$ 
of the ISM becomes
$$
p=C_S^2\mu\ ,\eqno(2.12)
$$
and similarly, the perturbed two-dimensional polytropic 
relation for the CRG is
$$
P=C_C^2\Sigma\ ,\eqno(2.13)
$$
where $C_S$ and $C_C$ are the polytropic sound speeds in the 
thermal ISM and in the CRG, respectively. In short, coplanar
MHD 
density wave perturbations in the ISM disk and the bulk of 
the CRG are coupled electromagnetically via the large-scale 
mean magnetic field.\footnote{$^4$}{Note that the large-scale 
dynamic coupling between the fluid stellar disk and the 
magnetized ISM gas disk is primarily gravitational (Jog \& 
Solomon 1984a, b; Lou \& Fan 1997, 1998b; Lou et al. 2001a).}

With the $\exp(i\omega t-im\theta)$ dependence implied for all 
MHD perturbation variables in equations $(2.3)-(2.13)$, where 
$\omega$ is the angular 
frequency in an inertial frame of reference and positive integer $m$ 
indicates the number of spiral arms, one can reduce 
equations (2.3)$-$(2.13) to 
$$
r(\omega-m\Omega)b_r=-mB_{\theta}v_r\ ,\eqno(2.14)
$$
$$
i\omega b_{\theta}={\partial (V_{\theta}b_r)\over\partial r}
-{\partial (B_{\theta}v_r)\over\partial r}\ ,\eqno(2.15)
$$
$$
b_{\theta}=-{i\over m}{\partial (rb_r)\over\partial r}\ ,\eqno(2.16)
$$
$$\eqalign{
i(\omega-m\Omega)v_r&-{2\Omega\mu_{\circ}v_{\theta}\over 
(\mu_{\circ}+\epsilon_{\circ})} 
-{2\Omega\epsilon_{\circ}u_{\theta}\over (\mu_{\circ}
+\epsilon_{\circ})}
+\int {dzB_{\theta}\over 4\pi r(\mu_{\circ}+\epsilon_{\circ})}
\bigg[{\partial (rb_{\theta})\over\partial r}+imb_r\bigg]
\cr &
={\partial\phi\over\partial r}
-{1\over (\mu_{\circ}+\epsilon_{\circ})}
{\partial (p+P)\over\partial r}+
{(\mu+\epsilon)\over (\mu_{\circ}+\epsilon_{\circ})^2}
{d(p_{\circ}+P_{\circ})\over dr}\ ,\cr}\eqno(2.17)
$$
$$
i(\omega-m\Omega)v_{\theta}+{1\over r}{d(rV_{\theta})\over dr}v_r
=-{im\over r}\bigg(\phi-{p\over\mu_{\circ}}\bigg)\ ,\eqno(2.18)
$$
$$
i(\omega-m\Omega)u_{\theta}+{1\over r}{d(rV_{\theta})\over dr}u_r
=-{im\over r}\bigg(\phi-{P\over\epsilon_{\circ}}\bigg)\ ,\eqno(2.19)
$$
$$
i(\omega-m\Omega)\mu+
{1\over r}{\partial (r\mu_{\circ}v_r)\over\partial r}
-{im\mu_{\circ}\over r}v_{\theta}=0\ ,\eqno(2.20)
$$
$$
i(\omega-m\Omega)\Sigma+
{1\over r}{\partial (r\Sigma_{\circ}u_r)\over\partial r}
-{im\Sigma_{\circ}\over r}u_{\theta}=0\ .\eqno(2.21)
$$
The three-dimensional Poisson equation (2.11) leads to an 
integral representation of $\phi$ in terms of $\mu$ (Shu 1970a; Lin 
\& Lau 1979; Shu et al. 2000; Galli et al. 2001; Lou \& Fan 2001), 
which can be solved exactly for potential-density pairs in special 
cases (e.g., logarithmic spirals; Kalnajs 1971). In the present 
context, we assume a sufficiently short radial wavelength and 
thus invoke the tight-winding or WKBJ approximation. In this regime, 
it is possible to establish a local differential relation between 
$\phi$ and $\mu$ for the magnetized thermal gas disk of ISM (Lin 
\& Shu 1964, 1966; Shu 1970b). For spiral galaxies such as M51 
and NGC 6946 of interest here, the tight-winding approximation is 
justifiable or, at least, should provide physically 
sensible results. We shall adopt this tight-winding approximation 
in our analytical analysis, as the solutions thus derived contain 
the information of MHD density waves in the presence of CRG fluid 
and of suprathermal MHD waves (see Parker 1965).

\hbox{ }
\vskip 0.4cm
\ctrline{3. DISPERSION RELATIONS OF SPIRAL MHD DENSITY WAVES}
\vskip 0.4cm

Parker first investigated suprathermal hydromagnetic waves using 
the MHD perturbation equations in a thermal gas and some very hot 
suprathermal gas, such as interstellar cosmic rays, in a uniform 
magnetic field. The results of his pioneer analysis provide 
useful insights on our problem at hand, although the formulation 
here involves additional effects of disk differential rotation, 
self-gravity, 
and a curved magnetic field in the cylindrical geometry. Parker 
showed that, besides an incompressible transverse Alfv\'en wave 
mode (see eq. [9] of Parker 1965), the conventional slow and fast 
hydromagnetic waves in the thermal gas are largely unaffected, 
except that there is a ``hole" in the phase diagram for the fast 
mode in the direction perpendicular to the magnetic field (see Fig. 1 
of Parker 1965; also see Appendix B here). The presence of this ``hole" 
in the fast mode is related to the fact that bulk motions of thermal 
and suprathermal gases are tied together electromagnetically in the 
direction transverse to the magnetic field. More importantly, there 
is an additional {\it suprathermal mode}, representing waves in the 
suprathermal gas traveling effectively with the speed of sound in 
the superthermal gas alone (Appendix B).

For spiral MHD density waves in a thin self-gravitating composite gas 
disk in rotation, we consider compressible MHD perturbations because 
mass density fluctuations are coupled to gravitational potential 
perturbation by Poisson's equation (2.11), so that a spiral 
gravitational potential field exerts a dynamical influence on 
compressible MHD perturbations. With a cylindrical geometry and in 
the presence of a thin differentially rotating disk, the analysis of 
MHD density waves becomes more tedious (Fan \& Lou 1996; Lou \& Fan 
1998; Lou et al. 2001a), yet the analogs of fast and slow MHD waves 
involving a suprathermal gas as described by Parker (1965) should 
exist on the ground of physics. We expect that for fast MHD density 
waves, thermal gas density and magnetic field perturbations remain 
more or less in phase in the tight-winding approximation, while 
thermal gas density and magnetic field perturbation enhancements 
remain significantly phase shifted for slow MHD density waves (a 
phase difference $\gsim\pi/2$ but not exactly $\pi$ owing to the 
disk rotation). The existence of a somewhat modified suprathermal 
mode is thus anticipated as well.

To derive the dispersion relations from equations
$(2.11)-(2.21)$ for the fast and slow spiral MHD density 
waves, we further take 
$$
\phi=\Phi(r)\exp\bigg[i\int^rk(s)ds\bigg]\ ,\eqno(3.1)
$$
where $k(r)$ is the radial wavenumber and $\Phi(r)$ is the 
slowly varying amplitude of the negative gravitational 
potential perturbation $\phi$ (Goldreich \& Tremaine 1978, 1979). 
One then has
$$
{d\phi\over dr}=ik(r)\phi+{\Phi'(r)\over\Phi(r)}\phi\ ,\eqno(3.2)
$$
where the prime over $\Phi(r)$ denotes a radial derivative. In 
the tight-winding regime of a large $k$, Poisson's equation 
(2.11) may be approximated by
$$
{1\over r^{1/2}}{\partial (r^{1/2}\phi)\over\partial r}\bigg|_{z=0}
\cong 2\pi Gi\hbox{ sgn}(k)\mu+{\cal O}\bigg[{\mu\over (kr)^2}\bigg]\ 
\eqno(3.3)
$$
which contains a fractional error on the order of ${\cal O}(kr)^{-2}$ 
(Shu 1970b; Goldreich \& Tremaine 1979), where $\hbox{sgn}(k)=+1$ for 
$k>0$ and $\hbox{sgn}(k)=-1$ for $k<0$. The cases of $k>0$ and $k<0$ 
correspond to leading and trailing spiral arms, respectively, and the 
integer $m>0$ gives the number of spiral arms. Equations 
(3.2) and (3.3) together give 
$$
\phi={4\pi r i G\hbox{ sgn}(k)\mu\over
2ikr+2r\Phi'/\Phi +1}\ .\eqno(3.4)
$$
A substitution of equation $(3.4)$ back into equation $(3.2)$ gives
$$
{d\phi\over dr}=\bigg[2ikr+{2r\Phi'\over\Phi}\bigg]
{2\pi i G\hbox{ sgn}(k)\mu\over 2ikr+2r\Phi'/\Phi +1}
\defeq {\cal F}_1\ \mu\ , \eqno(3.5)
$$
which defines the complex coefficient ${\cal F}_1$. From 
equations (2.12), (2.18), (2.21) and (3.4), we can derive 
an equation of $\mu$ in terms of $v_r$ and $dv_r/dr$ 
$$\eqalign{
\mu=&{im\kappa^2\mu_{\circ}v_r/(2\Omega r)+i(\omega-m\Omega)r^{-1}
d(r\mu_{\circ}v_r)/dr \over (\omega-m\Omega)^2-m^2C_S^2/r^2+(m^2/r^2)
4\pi ri G\hbox{ sgn}(k)\mu_{\circ}/(2ikr+2r\Phi'/\Phi +1)}\cr &
\qquad\qquad\qquad \defeq {\cal A}v_r+{\cal B}{dv_r\over dr}\ ,\cr}
\eqno(3.6)
$$
which defines the two complex coefficients ${\cal A}$ and 
${\cal B}$. From equations (2.13) and (2.19) for perturbations 
in the CRG, one derives 
$$
u_{\theta}={iu_r\over r(\omega-m\Omega)}{d(rV_{\theta})\over dr}
+{mC_C^2\Sigma\over r(\omega-m\Omega)\epsilon_{\circ}}\ .\eqno(3.7)
$$
From mass conservation (2.21) for perturbations of the CRG, one obtains
$$
\Sigma={im\Sigma_{\circ}\kappa^2u_r/(2\Omega r)+
i(\omega-m\Omega)r^{-1}d(r\Sigma_{\circ}u_r)/dr\over 
(\omega-m\Omega)^2-m^2C_C^2\Sigma_{\circ}/(r^2\epsilon_{\circ})}\ ,
\eqno(3.8)
$$
where $\kappa^2=(2\Omega/r)d(rV_{\theta})/dr$ defines the epicyclic 
frequency $\kappa$ of the disk. From the mass conservation (2.20) 
for the thermal ISM, one has
$$
v_{\theta}={r(\omega-m\Omega)\mu\over m\mu_{\circ}}
-{i\over m\mu_{\circ}}{d(r\mu_{\circ}v_r)\over dr}\ .\eqno(3.9)
$$
The radial induction equation (2.14) 
may be arranged into the form of
$$
b_r=-{mB_{\theta}v_r\over r(\omega-m\Omega)}\ ,\eqno(3.10)
$$
while the divergence-free condition 
(2.16) of $\vec b$ may be written as
$$
b_{\theta}=i{d\over dr}\bigg[{B_{\theta}v_r
\over (\omega-m\Omega)}\bigg] \ \eqno(3.11)
$$
by using equation (3.10). A straightforward substitution of these 
equations into the radial momentum equation (2.17) leads to the 
following lengthy second-order ordinary differential equation (ODE)
$$\eqalign{
&i(\omega-m\Omega )v_r -{2\Omega\mu_{\circ}\over 
(\mu_{\circ}+\epsilon_{\circ})}\bigg[{r(\omega-m\Omega )\over
m\mu_{\circ} }\bigg({\cal A}v_r+{\cal B}{dv_r\over dr}\bigg)
-{i\over m\mu_{\circ}}{d(r\mu_{\circ}v_r)\over dr}\bigg]\cr &
-{2\Omega\epsilon_{\circ}\over (\mu_{\circ}+\epsilon_{\circ})}
\bigg\lbrace 
{mC_C^2\Sigma_{\circ}\over r(\omega-m\Omega )\epsilon_{\circ}}
\bigg[{im\kappa^2u_r/(2\Omega r)+i(\omega-m\Omega)d(r\Sigma_{\circ}u_r)
/dr/(r\Sigma_{\circ})\over (\omega-m\Omega)^2-m^2C_C^2\Sigma_{\circ}
/(r^2\epsilon_{\circ})}\bigg]\cr &
+{i\kappa^2 u_r\over 2\Omega (\omega-m\Omega)} \bigg\rbrace
-{im^2 C_A^2v_r\over r^2(\omega-m\Omega)}
+iC_A^2{d\over dr}\bigg\lbrace r{d\over dr}\bigg[
{v_r\over r(\omega-m\Omega)}\bigg]\bigg\rbrace \cr &
={\cal F}_1\bigg({\cal A}v_r+{\cal B}{dv_r\over dr}\bigg)
-{1\over (\mu_{\circ}+\epsilon_{\circ})}
{d\over dr}\bigg[C_S^2\bigg({\cal A}v_r
+{\cal B}{dv_r\over dr}\bigg)\bigg]\cr &
-{1\over (\mu_{\circ}+\epsilon_{\circ})}
{d\over dr}\bigg\lbrace {C_C^2\Sigma_{\circ}
[im\kappa^2u_r/(2\Omega r)+i(\omega-m\Omega)d(r\Sigma_{\circ}u_r)
/dr/(r\Sigma_{\circ})]\over (\omega-m\Omega)^2-m^2C_C^2\Sigma_{\circ}
/(r^2\epsilon_{\circ})}\bigg\rbrace \cr &
+\bigg({\cal A}v_r+{\cal B}{dv_r\over dr}\bigg)
{1\over (\mu_{\circ}+\epsilon_{\circ})^2}{d(p_{\circ}+P_{\circ})
\over dr} \cr &
+{\Gamma C_C^2\Sigma_{\circ}/[(\Gamma-1)c^2]\over 
(\mu_{\circ}+\epsilon_{\circ})^2}{d(p_{\circ}+P_{\circ})\over dr}
{[im\kappa^2u_r/(2\Omega r)+i(\omega-m\Omega)d(r\Sigma_{\circ}u_r)
/dr/(r\Sigma_{\circ})]\over (\omega-m\Omega)^2-m^2C_C^2\Sigma_{\circ}
/(r^2\epsilon_{\circ})},\cr}\eqno(3.12)
$$
where $C_A^2\equiv\int dzB_{\theta}^2/
[4\pi (\mu_{\circ}+\epsilon_{\circ})]$ defines the Alfv\'en speed 
$C_A$ in the composite disk, and the complex coefficients ${\cal F}_1$, 
${\cal A}$, and ${\cal B}$ have already been defined earlier by 
equations (3.5) and (3.6) (see Appendix A for details). With 
$v_r=u_r$, equation (3.12) contains the information of fast and 
slow spiral MHD density waves as well as the suprathermal MHD mode. 

\hbox{ }
\vskip 0.4cm
\line{3.1 SLOW SPIRAL MHD DENSITY WAVES \hfill}
\vskip 0.4cm

We have previously (Fan \& Lou 1996; Lou \& Fan 1998a) derived, 
in the tight-winding approximation, the dispersion relation for 
slow MHD density waves 
under the conditions that $\Omega\equiv V_{\theta}/r\oforder$ constant, 
$\omega-m\Omega\oforder mC_A/r$, $C_S\oforder C_A$, $\Omega/C_A\sim k$, 
and $kr\grgr 1$. Here, we require in addition that 
$C_C\gg C_S$ and $\Sigma_{\circ}/\mu_{\circ}\ll 1$. On the basis of 
Parker's analysis (Parker 1965), we expect that the basic features 
of slow MHD density waves should remain more or less intact and the 
associated CRG density fluctuation should be fairly weak. We 
set out to confirm these expectations by an analysis of equation 
(3.12) (see Appendix A) using the assumptions stated above. 

To the leading order of the tight-winding approximation with
$kr\grgr 1$ for slow MHD density waves, the three complex 
coefficients ${\cal C}_1$, ${\cal C}_2$ and ${\cal C}_3$ in 
equation $(A1)$ are approximately given by
$$\eqalign{
{\cal C}_1={iC_A^2\over (\omega-m\Omega)}&+{C_S^2\mu_{\circ}\over
(\mu_{\circ}+\epsilon_{\circ})}{i(\omega-m\Omega)
\over (\omega-m\Omega)^2-m^2\Delta/r^2} \cr &
+{i(\omega-m\Omega) C_C^2\Sigma_{\circ}\over (\mu_{\circ}+\epsilon_{\circ})
[(\omega-m\Omega)^2-m^2C_C^2\Sigma_{\circ}
/(r^2\epsilon_{\circ})]}\ \ , \cr }\eqno(3.13)
$$
$$\eqalign{
{\cal C}_2=&-{2\Omega\mu_{\circ}\over (\mu_{\circ}+\epsilon_{\circ})}
\bigg[{ir(\omega-m\Omega)^2\over m[(\omega-m\Omega)^2-m^2\Delta/r^2]}
-{ir\over m}\bigg] \cr &
-{2im\Omega C_C^2\Sigma_{\circ}\over r(\mu_{\circ}+\epsilon_{\circ})
[(\omega-m\Omega)^2-m^2C_C^2\Sigma_{\circ}/(r^2\epsilon_{\circ})]}\cr &
-ik {2\pi G\mu_{\circ}\over |k|}
{i(\omega-m\Omega)\over (\omega-m\Omega)^2-m^2\Delta/r^2} \cr &
+ {C_C^2\Sigma_{\circ}im\kappa^2/(2\Omega r)
\over (\mu_{\circ}+\epsilon_{\circ})
[(\omega-m\Omega)^2-m^2C_C^2\Sigma_{\circ}/(r^2\epsilon_{\circ})]} \cr &
+{C_S^2\mu_{\circ}\over (\mu_{\circ}+\epsilon_{\circ})} 
{im\kappa^2/(2\Omega r)\over (\omega-m\Omega)^2-m^2\Delta/r^2}
\ \ ,\cr}\eqno(3.14)
$$
and
$$\eqalign{
{\cal C}_3=- &{\mu_{\circ}\over
(\mu_{\circ}+\epsilon_{\circ})}{i(\omega-m\Omega )
\kappa^2\over (\omega-m\Omega)^2-m^2\Delta/r^2} 
-{2\Omega\epsilon_{\circ}\over (\mu_{\circ}+\epsilon_{\circ})}
\bigg [{i\kappa^2\over 2\Omega (\omega-m\Omega)} \cr &
+{mC_C^2\Sigma_{\circ}\over r(\omega-m\Omega )\epsilon_{\circ}}
{im\kappa^2/(2\Omega r)
\over (\omega-m\Omega)^2-m^2C_C^2\Sigma_{\circ}/
(r^2\epsilon_{\circ})}\bigg ]\cr &
-ik{2\pi G\mu_{\circ}\over |k|}
{im\kappa^2/(2\Omega r)\over (\omega-m\Omega)^2-m^2\Delta/r^2}
\ \cr}\eqno(3.15)
$$
separately with $\Delta\equiv C_S^2-2\pi G\mu_{\circ}/|k|$
(see Appendix A). 
Ignoring higher-order terms, the dispersion relation of 
slow MHD density waves comes out to be
$$\eqalign{
(\omega-m\Omega)^2\cong {k^2 C_A^2m^2\Delta\over r^2}
\bigg\lbrace
k^2\bigg[C_A^2&+{C_S^2\mu_{\circ}\over (\mu_{\circ}+\epsilon_{\circ})}
+{\epsilon_{\circ}\Delta\over (\mu_{\circ}+\epsilon_{\circ})}
-{2\pi G\mu_{\circ}\over |k|}\bigg] \cr &
+\kappa^2\bigg[{\mu_{\circ}\over (\mu_{\circ}+\epsilon_{\circ})}
+{\epsilon_{\circ}^2\Delta\over C_C^2\Sigma_{\circ}
(\mu_{\circ}+\epsilon_{\circ})}\bigg] \bigg\rbrace^{-1}\ .\cr}
\eqno(3.16)
$$
Because $\mu_{\circ}\grgr \epsilon_{\circ}\hbox{ or }\Sigma_{\circ}$,
the slow MHD density wave is indeed not very much affected by the presence
of a tenuous CRG (Fan \& Lou 1996; Lou \& Fan 1998a, 2001). As 
expected, perturbation enhancements of surface mass density $\mu$ of the 
thermal ISM gas and parallel magnetic field $b_{\theta}$ are significantly 
phase shifted with a phase difference $\gsim\pi/2$. The particular aspect 
of interest here is the phase relation and magnitude of CRG density 
fluctuation $\Sigma$ relative to thermal gas density fluctuation $\mu$. 
By equations (3.6) and (3.8), we have for large $kr$
$$
{\Sigma/\Sigma_{\circ}\over\mu/\mu_{\circ}}
\cong {(\omega-m\Omega)^2-m^2\Delta/r^2\over
(\omega-m\Omega)^2-m^2C_C^2\Sigma_{\circ}/(r^2\epsilon_{\circ})}\ .
\eqno(3.17)
$$
For neutral or stable slow MHD density waves with $\Delta>0$,
the inequality $(\omega-m\Omega)^2 < m^2\Delta/r^2
\lsls m^2C_C^2\Sigma_{\circ}/(r^2\epsilon_{\circ})$ holds
such that $\Sigma/\Sigma_{\circ}$ and $\mu/\mu_{\circ}$ are 
in phase with 
$$
{\Sigma\over\Sigma_{\circ}}\oforder {\Delta\epsilon_{\circ}
\over C_C^2\Sigma_{\circ}}{\mu\over\mu_{\circ}}\ .\eqno(3.18)
$$
As $\epsilon_{\circ}$ and $\Sigma_{\circ}$ are on the same order 
of magnitude and $\Delta\ll C_C^2$, the relative fluctuation 
$\Sigma/\Sigma_{\circ}$ of CRG density is much less than 
the relative fluctuation $\mu/\mu_{\circ}$ of thermal gas density.
Therefore, the main contribution to the large-scale variation of 
synchrotron radio-continuum emission comes from the enhancement of magnetic 
field. In other words, the significant phase difference ($\gsim\pi/2$) 
between the enhancements of CRG density and parallel magnetic 
field would not reduce the contrast of synchrotron radio-continuum
emission associated with slow MHD density waves in a significant manner.

As the presence of tenuous CRG does not influence the 
propagation of slow MHD density waves in a significant manner, 
it follows that the dispersion relation for slow MHD density waves 
in a disk with a strong differential rotation should remain more or 
less the same as derived earlier (Lou \& Fan 1998a). 

\hbox{ }
\vskip 0.4cm
\line{3.2 FAST SPIRAL MHD DENSITY WAVES \hfill}
\vskip 0.4cm

To derive the dispersion relation of fast MHD density 
waves in the presence of a CRG, we take that 
$(\omega-m\Omega)\oforder (C_S^2+C_A^2)^{1/2}k$, 
$C_S\oforder C_A$, $\Omega/C_A\oforder k$, $kr\grgr 1$, 
$C_C\gg C_S$, $\Sigma_{\circ}\ll\mu_{\circ}$, and 
$m/(kr)\grgr C_S/C_C$, where the last inequality guarantees 
that the direction of wave propagation be sufficiently away 
from the radial direction. It is then straightforward 
to show (see Appendix A) that
$$\eqalign{
{\cal C}_1= {iC_A^2\over (\omega-m\Omega)}&+{C_S^2\mu_{\circ}\over
(\mu_{\circ}+\epsilon_{\circ})}{i(\omega-m\Omega)\over
(\omega-m\Omega)^2-m^2\Delta/r^2} \cr &
+{i(\omega-m\Omega) C_C^2\Sigma_{\circ}\over
(\mu_{\circ}+\epsilon_{\circ})
[(\omega-m\Omega)^2-m^2C_C^2\Sigma_{\circ}
/(r^2\epsilon_{\circ})]}\ ,\cr}\eqno(3.19)
$$
where the third term on the right-hand 
side due to the CRG is small, 
$$
{\cal C}_2=k{2\pi G\mu_{\circ}\over |k|}
{(\omega-m\Omega)\over
(\omega-m\Omega)^2-m^2\Delta/r^2}\ ,\eqno(3.20)
$$
and
$$\eqalign{
{\cal C}_3=i(\omega &-m\Omega)-{i\mu_{\circ}\over 
(\mu_{\circ}+\epsilon_{\circ})}{(\omega-m\Omega )
\kappa^2\over (\omega-m\Omega)^2-m^2\Delta/r^2 }
-{\epsilon_{\circ}i\kappa^2\over (\mu_{\circ}+\epsilon_{\circ})
(\omega-m\Omega)} \cr &
-{2\Omega\over (\mu_{\circ}+\epsilon_{\circ})}
{mC_C^2\Sigma_{\circ}\over r(\omega-m\Omega )}
{im\kappa^2/(2\Omega r)+i(\omega-m\Omega)d\ln (r\Sigma_{\circ})
/dr\over (\omega-m\Omega)^2-m^2C_C^2\Sigma_{\circ}/
(r^2\epsilon_{\circ})} \ \cr}\eqno(3.21)
$$
in equation $(A1)$. The dispersion relation of 
fast spiral MHD density waves is thus given by
$$
(\omega-m\Omega)^2\cong {\mu_{\circ}\kappa^2\over 
(\mu_{\circ}+\epsilon_{\circ})}
+k^2\bigg[C_A^2+{C_S^2\mu_{\circ}\over 
(\mu_{\circ}+\epsilon_{\circ})}
-{2\pi G\mu_{\circ}\over |k|}\bigg]\ .\eqno(3.22)
$$
As expected, the presence of CRG does not affect very much the 
propagation of fast MHD density waves as long as $m/(kr)\grgr C_S/C_C$. 
This inequality guarantees that a propagation of fast MHD density waves 
is sufficiently away from the radial direction that is perpendicular to 
the background magnetic field (see Appendix B). One can readily show 
that the perturbation enhancements of thermal gas density and parallel 
magnetic field are largely in phase to the leading order of large $kr$. 
From equation (3.17),
it is clear that the perturbation enhancements of thermal ISM 
density and CRG density are out of phase (i.e., a phase difference 
of $\oforder\pi$). Furthermore,
$$
{\Sigma/\Sigma_{\circ}\over\mu/\mu_{\circ}}
\oforder -{k^2r^2 C_A^2\over m^2C_C^2}\ ,\eqno(3.23)
$$
which gives a much smaller $\Sigma/\Sigma_{\circ}$ relative to
$\mu/\mu_{\circ}$. Nevertheless, in comparison to the case of slow 
MHD density waves, $\Sigma/\Sigma_{\circ}$ is larger by a factor 
$k^2r^2/m^2$ in the case of fast MHD density waves. Although 
perturbation enhancements of CRG density and parallel 
magnetic field are out of phase, the synchrotron radio-continuum
emission is dominantly determined by the perturbation 
enhancement of large-scale parallel magnetic field.

\hbox{ }
\vskip 0.4cm
\line{3.3 THE SUPRATHERMAL MODE \hfill}
\vskip 0.4cm

To derive the dispersion relation of the suprathermal 
mode, we take that $(\omega-m\Omega)\oforder mC_C/r$, 
$C_S\oforder C_A$, $\Omega/C_A\oforder k$, $C_C\gg C_S$, 
$\Sigma_{\circ}\ll\mu_{\circ}$, and $kr\grgr 1$. 
It follows that in equation $(A1)$
$$\eqalign{
{\cal C}_1= {iC_A^2\over (\omega-m\Omega)}&+{C_S^2\mu_{\circ}\over
(\mu_{\circ}+\epsilon_{\circ})}{i(\omega-m\Omega)\over
(\omega-m\Omega)^2-m^2\Delta/r^2} \cr &
+{i(\omega-m\Omega) C_C^2\Sigma_{\circ}\over
(\mu_{\circ}+\epsilon_{\circ})
[(\omega-m\Omega)^2-m^2C_C^2\Sigma_{\circ}
/(r^2\epsilon_{\circ})]}\ ,\cr}\eqno(3.24)
$$
$$
{\cal C}_2=k{2\pi G\mu_{\circ}\over |k|}
{(\omega-m\Omega)\over
(\omega-m\Omega)^2-m^2\Delta/r^2}\ ,\eqno(3.25)
$$
and
$$
{\cal C}_3=i(\omega -m\Omega)-{i\mu_{\circ}\over
(\mu_{\circ}+\epsilon_{\circ})}{(\omega-m\Omega )
\kappa^2\over (\omega-m\Omega)^2-m^2\Delta/r^2 }\ \eqno(3.26)
$$
(see Appendix A). By dropping several 
higher-order terms, we arrive at 
$$\eqalign{
(\omega&-m\Omega)^4-\bigg\lbrace {\mu_{\circ}\kappa^2
\over (\mu_{\circ}+\epsilon_{\circ})}
+k^2\bigg[C_A^2+{C_S^2\mu_{\circ}\over (\mu_{\circ}+\epsilon_{\circ})}
-{2\pi G\mu_{\circ}\over |k|}\bigg] \cr &
+{k^2C_C^2\Sigma_{\circ}\over (\mu_{\circ}+\epsilon_{\circ})}
+{m^2C_C^2\Sigma_{\circ}\over r^2\epsilon_{\circ}}\bigg\rbrace
(\omega-m\Omega)^2 \cr &
+\bigg\lbrace {\mu_{\circ}\kappa^2
\over (\mu_{\circ}+\epsilon_{\circ})}
+k^2\bigg[C_A^2+{C_S^2\mu_{\circ}\over (\mu_{\circ}+\epsilon_{\circ})}
-{2\pi G\mu_{\circ}\over |k|}\bigg]\bigg\rbrace
{m^2C_C^2\Sigma_{\circ}\over r^2\epsilon_{\circ}}\cong 0\ , \cr}
\eqno(3.27)
$$
which contains both the suprathermal wave mode and the fast MHD density
wave mode. If the term $(\omega-m\Omega)^2k^2C_C^2\Sigma_{\circ}
/(\mu_{\circ}+\epsilon_{\circ})$ were absent, then relation (3.27)
may be readily expressed as a multiplication of two simple factors.
To the present level of approximation, the dispersion relation for
the suprathermal mode is approximately given by
$$
(\omega-m\Omega)^2\cong {m^2C_C^2\Sigma_{\circ}\over r^2\epsilon_{\circ}}
+{k^2C_C^2\Sigma_{\circ}\over (\mu_{\circ}+\epsilon_{\circ})}
+{\mu_{\circ}\kappa^2
\over (\mu_{\circ}+\epsilon_{\circ})}
+k^2\bigg[C_A^2+{C_S^2\mu_{\circ}\over (\mu_{\circ}+\epsilon_{\circ})}
-{2\pi G\mu_{\circ}\over |k|}\bigg]\ .\eqno(3.28)
$$
Again, from the relation (3.17),
it is seen that $\Sigma/\Sigma_{\circ}$ 
and $\mu/\mu_{\circ}$ are in phase with 
$$
{\Sigma\over\Sigma_{\circ}}\oforder 
{C_C^2m^2\over C_S^2k^2r^2}{\mu\over\mu_{\circ}}\ ,\eqno(3.29)
$$
where the magnitude of $\Sigma/\Sigma_{\circ}$ may be 
comparable to or larger than that of $\mu/\mu_{\circ}$ for the 
suprathermal mode.

The velocity perturbations of the suprathermal mode can 
be derived from the following relationship 
$$\eqalign{
u_{\theta}=
{mC_C^2\Sigma_{\circ}\over r(\omega-m\Omega)\epsilon_{\circ}}
&{im\kappa^2u_r/(2\Omega r)+i(\omega-m\Omega)
[u_rd\ln (r\Sigma_{\circ})/dr+du_r/dr ]
\over (\omega-m\Omega)^2
-m^2C_C^2\Sigma_{\circ}/(r^2\epsilon_{\circ})}\cr &
+{i\kappa^2u_r\over 2\Omega (\omega-m\Omega)}\ .\cr }\eqno(3.30)
$$
For an order-of-magnitude estimates, we have
$$
{u_{\theta}\over u_r}\oforder -{mC_C^2\Sigma_{\circ}
\over kr C_S^2\epsilon_{\circ}}\ .\eqno(3.31)
$$
For $u_r\oforder 1\hbox{ km s}^{-1}$, the azimuthal velocity 
perturbation $u_{\theta}$ could be as large as $\sim 10^4
\hbox{ km s}^{-1}$. For $u_{\theta}\gsim C_A$, one should be 
seriously concerned with streaming instabilities along the magnetic 
field. Parker (1969) pointed out the possibility of Landau damping 
for the suprathermal mode in the CRG.

\hbox{ }
\vskip 0.4cm
\ctrline{4. APPLICATIONS TO SPIRAL GALAXIES}
\vskip 0.4cm

To apply the results of our analyses to large-scale spiral structures 
of gas-rich disk galaxies seen in synchrotron radio-continuum emissions, 
one needs to have an idea for the plausible state of CRG in a 
typical magnetized disk galaxy. If the background CRG as a 
whole (either a fat disk or an oblated spheroid) largely corotates with 
the stellar and magnetized gas disks while retains large-scale 
axisymmetry, then our theoretical results indicate the following. Firstly, 
in the presence CRG density fluctuations on large spatial scales, the 
suprathermal mode propagates essentially with the cosmic-ray sound speed 
(always less than but very close to the speed of light $c$) more or less 
along magnetic field lines to smooth out CRG fluctuations. In this case, 
the large-scale distribution of CRG is likely to vary in the 
radial direction due to the presence of a mean azimuthal galactic 
magnetic field. Secondly, CRG density fluctuations associated 
with slow MHD density waves are extremely weak as compared to thermal 
gas density fluctuations. While the perturbation enhancements of 
CRG density and parallel magnetic field are significantly 
phase shifted (i.e., $\gsim\pi/2$), the enhancement in synchrotron 
radio-continuum emission is mainly determined by the enhancement of 
parallel magnetic field. Thirdly, CRG density fluctuations 
associated with fast MHD density waves are also sufficiently weaker 
than thermal gas density fluctuations to warrant that the enhancement 
of synchrotron radio-continuum emission basically goes with the 
enhancement of parallel magnetic field; the latter anti-correlates 
with the minute enhancement of CRG density fluctuation.

For an actual spiral galaxy, several complications and uncertainties
arise. The key question ties to the actual distribution of 
relativistically hot CRG in a typical disk spiral galaxy. We further 
elaborate several relevant aspects below.

1. It is known observationally that the total luminosities of infrared 
and nonthermal radio-continuum emissions from spiral galaxies are 
tightly correlated (Dickey \& Salpeter 1984; Helou et al. 1985; 
Helou \& Bicay 1993). This empirical fact leads to the important
notion that somehow the formation of massive stars and the production 
of relativistic cosmic rays are correlated through a chain of complex 
physical processes yet to be explored and understood. For example, an 
active region of massive star formation may produce, perhaps in the 
statistical sense, more explosions of type II supernovae that are 
favored sources of cosmic ray production (up to energy of 
$\sim 10^{15}$ eV). It is also plausible that MHD termination shocks 
confining fast stellar winds from numerous young O, B stars sustain 
the supply of galactic cosmic rays in a significant manner. It thus 
seems sensible that on large spatial scales, optical spiral gas arms 
of relatively high densities should contain more sources for cosmic 
ray production.

2. Interstellar cloud complexes, where young massive stars are 
continuously forming on smaller scales, are generally thought to 
be initially triggered by Parker instability (i.e., magnetic 
Rayleigh-Taylor instability, see Parker 1966, 1969; Shu 1974; 
Mouschovias 1996) which involves downward slide of condensed 
thermal gas into magnetic valleys and upward expansions of 
magnetic flux tubes further inflated by CRG in the dimension 
perpendicular to the galactic disk. To account for such processes, 
one must allow for MHD fluctuations perpendicular to a galactic 
plane. The development of 
Parker instability disturbs galactic magnetic field on horizontal 
size scales of $\oforder 1\hbox{ kpc}$ or so. Subsequent activities 
of star formation can lead to intensified 
but disordered magnetic fields on smaller spatial scales via ISM 
turbulence. Therefore, the total magnetic field strength may 
increase with star formation activities in high-density gas arms.
As enhanced magnetic fields in this manner tend to be disordered, 
the degree of polarization in radio-continuum emissions should be 
correspondingly low. Therefore for fast MHD density waves with 
in-phase gas density and magnetic field enhancements, manifestations 
of total and polarized radio-continuum emissions are in competition.
That is, along a high-density gas arm with strong star formation 
activities, the total radio-continuum emission would be very strong 
whereas the degree of polarization would become weak. For 
moderate star formation activities along high-density gas arms, 
total emissions are generally stronger than polarized emissions 
but one may still detect degrees of polarizations. For both M51 
(Neininger 1992) and NGC 2997 (Han et al. 1999), polarized radio
intensity peaks closely follow the inner edges of optical spiral 
arms and coincide with narrow dark dust lanes. For M31 (Beck et 
al. 1980), magnetic ``torus" coincide with the ``ring" (for the 
controversy of ``ring" or tight-winding spiral arms for M31, the 
reader is referred to references in Berkhuijsen et al. 2000). On 
the other hand, for slow MHD density waves with phase-shifted gas 
density and magnetic field enhancements, polarized radio intensity 
arms are much less affected by star formation activities along 
high-density gas arms where total radio emissions still remain to 
be strong. In terms of overall structural patterns, NGC 6946 fits 
this scenario well (Lou \& Fan 2002).

3. Observations indicate that large-scale galactic magnetic field 
normally lies more or less in the disk plane of a spiral galaxy. Within 
the disk of a nearly face-on spiral galaxy (e.g., M51, NGC 2997 etc), 
polarized radio-continuum emissions reveal a global swirl-like 
pattern in the same sense of optical spiral arms (Neininger 1992; 
Berkhuijsen et al. 1997; Han et al. 1999).\footnote{$^5$}{At longer 
radio wavelengths, the effect of Faraday rotation gives rise to 
more distorted or irregular swirl-like patterns (e.g.,
Berkhuijsen et al. 1997).} 
The usual impression that magnetic field remains aligned and 
{\it connected} along spiral arms would be difficult to reconcile 
with the existence of a strong disk differential rotation (Lynden-Bell 
1966; Roberts \& Yuan 1970). Disk differential rotation 
would wrap up a connected magnetic field line along a spiral arm in 
a few turns of galactic rotation. Similar to the winding dilemma of 
material spiral arms (Goldreich \& Lynden-Bell 1965; Lin \& Shu 1966), 
this winding dilemma for the magnetic field may be resolved by 
considering large-scale azimuthal galactic magnetic field strongly 
distorted by galactic MHD density waves in a systematic manner; it 
is plausible that enhanced magnetic field locally orients 
{\it roughly} along a spiral arm but is not actually connected along 
the spiral arm. In this scenario, magnetic spiral arms appear as a 
result of the collective behavior of phase-organized, distorted rings 
of magnetic field and gas stream lines\footnote{$^6$}{This is very much 
like the kinematic description of density waves in a stellar disk
by visualizing the collective behavior of phase-organized, distorted 
rings of mean stellar orbits as suggested by Kalnajs (1973).}. 
We note that M51 seen in linearly polarized optical emissions shows 
a more circular or circumferential pattern of magnetic field 
configuration (e.g., Scarrott, Ward-Thompson \& Warren-Smith 1987).

4. For the scenario of enhanced magnetic field aligned and connected 
along a high-density spiral gas arm, cosmic rays produced by star 
formation and type II supernovae would be somewhat confined to retain 
relatively high number density along the arm. While for the scenario 
of distorted rings of magnetic field associated with MHD density waves, 
cosmic rays produced in high gas density spiral arms can redistribute 
themselves more readily along distorted rings of magnetic field, 
maintaining a large-scale radial gradient across rings of magnetic 
field at the same time. For a mature late-type spiral galaxy, the 
level of cosmic rays has probably reached the maximum that a galaxy 
can contain and remains more or less steady (i.e., leakage and 
decays are globally balanced by production of cosmic rays). In the 
azimuthal direction, a relatively high concentration of cosmic rays 
can be smoothed out via propagation of the suprathermal mode (Parker 
1965, 1969), by various diffusion processes (see Cesarsky 1980) and 
by scattering processes.

5. It is the more common situation that the spiral pattern of the total 
(random and organized magnetic fields plus cosmic rays) synchrotron 
radio-continuum emission more or less coincides with the optical 
spiral pattern in nearby spiral galaxies. Typically, a spiral arm 
of total synchrotron radio-continuum emission is broader and may 
extend further outside optical arms. 
This demonstrates the ability of cosmic rays to diffuse across 
magnetic field and to infiltrate a much larger spatial volume than 
their original sources of production. Qualitatively, one 
tends to interpret the optical-radio spiral pattern correlation in 
terms of somewhat high concentration of cosmic rays and enhanced 
random magnetic fields along a spiral arm of high thermal gas 
density where stars are actively forming. As large-scale organized 
magnetic field may not be completely disrupted along star-forming 
regions, coherent structures in polarized radio-continuum emissions 
can be still detected.

6. In addition to the fact that {\it total} luminosities of infrared 
and synchrotron radio-continuum emission from spiral galaxies are 
tightly correlated, new evidence is forthcoming for the infrared-radio 
correlation in spiral structural patterns within disk galaxies 
(e.g., M31, see
references in Berkhuijsen et al. 2000 and Lou \& Fan 2000a, b), which 
is quite similar to optical-radio correlation in spiral patterns. With 
increasing sensitivity and angular resolutions of infrared observations, 
we expect more examples of detailed spiral arm correlations. This trend 
of infrared-radio correlation in spiral patterns would be consistent 
with the general perception of spiral MHD density waves, even though 
the chain of physical processes on various scales that lead to such 
correlated radiative manifestations remain to be thoroughly understood.

7. In identifying fast MHD density waves with large-scale optical 
and radio spiral patterns in M51 (Fan \& Lou 1996; Lou \& Fan 1998a), 
we follow a few clues. Firstly, both polarized intensity maps and 
degree-of-polarization maps for synchrotron radio-continuum emissions 
from M51 (Neininger 1992; Berkhuijsen et al. 1997) reveal a global 
spiral pattern well correlated with the optical spiral pattern. 
This is the strongest evidence that large-scale galactic magnetic 
fields are involved in the dynamics of density waves. Secondly, both 
polarized intensity and degree of polarization are usually strong along
optical spiral arms where thermal gas concentration is relatively high.
The coincident enhancement of thermal gas density and magnetic field is 
expected from the basic property of fast MHD density waves. This implies
that the process of star formation does not completely disrupt large-scale
galactic magnetic field, although one cannot tell immediately
whether CRG density 
is much higher in spiral arms than in interarm regions. Even if CRG 
density distributes rather evenly in the azimuthal direction, it 
is still expected to see enhanced polarized and total radio-continuum 
intensities. Finally, from the perspective of self excitation and 
maintenance of MHD density waves\footnote{$^7$}{The presence of a 
companion (NGC 5195) around M51 (NGC 5194) is probably responsible only 
for the distortion in the {\it outer} spiral pattern as a result of 
tidal interaction (Toomre \& Toomre 1972; Elmegreen et al. 1989); the 
{\it inner} spiral pattern may be self-sustained (Rix \& Rieke 1993; 
Lin 1996).\vskip -16pt}, fast MHD density waves are preferentially 
swing amplified in a disk of strong differential rotation (Fan \& 
Lou 1997) which is the case for M51 as well as for NGC 2997.

8. In identifying slow MHD density waves with large-scale optical and 
radio spiral patterns in the spiral galaxy NGC 6946 (Beck \& Hoernes 
1996; Beck et al. 1996; Fan \& Lou 1996; Ferguson et al. 1998; Frick 
et al. 2000; Lou \& Fan 1998a, 2002), we would like to clarify 
a few points. Firstly, although somewhat fuzzy, large-scale radio 
spiral pattern in NGC 6946 again indicates the involvement of magnetic 
field in global density waves. The {\it total} synchrotron 
radio-continuum emission along optical spiral arms is strong (see Fig. 3 
of Beck \& Hoernes 1996). This can be explained, as usual, in terms of 
the enhancement of small-scale random magnetic fields and cosmic ray 
production along 
zones of relatively high thermal gas concentration where stars are 
continuously borne. Secondly, intensity and degree of polarized radio 
emission are sufficiently strong to reveal global magnetic arms interlaced 
with optical spiral arms\footnote{$^8$}{Even in the interarm regions, total 
radio-continuum emission is several times stronger than the peak polarized 
radio-continuum intensity. However, an interarm bump in total radio emission 
can be attributed to polarized intensity peak.\vskip -16pt} 
(see Fig. 2 of Beck \& 
Hoernes 1996 and Frick et al. 2000). For slow MHD density waves (Fan \& 
Lou 1996; Lou \& Fan 1998a), the enhancements of parallel magnetic field 
and thermal gas density are significantly phase shifted (i.e., 
$\gsim\pi/2$), and the {\it associated} fluctuation of CRG density is so 
weak as can be neglected. It is therefore natural to interpret the 
prominent magnetic arms in polarized radio-continuum intensity in terms 
of enhancements of large-scale organized magnetic field, associated with 
slow MHD density waves, submerged in an omnipresent CRG. In this scenario, 
magnetic arms are interlaced with the optical spiral arms as a result of 
significant phase shift between enhancements of parallel magnetic field 
and thermal gas density. The fact that the polarized radio intensity is 
extremely low along optical arms (or total radio intensity arms; see 
Fig. 3 of Beck \& Hoernes 1996) may be due to a nearly complete disruption
of an already weakened large-scale magnetic field or due to an almost flat 
distribution of CRG along the azimuthal direction\footnote{$^9$}
{That is, CRG density enhancement across star-forming optical 
arms may be smoothed out sufficiently fast such that the distribution 
remains more or less uniform in the azimuthal direction along magnetic 
field lines.\vskip -16pt} or both. At least, the CRG density 
cannot be too much higher across optical arms than in interarm regions; 
otherwise, the presence of a weak organized large-scale magnetic field 
may lead to a polarized radio-continuum intensity peak within optical 
arms as well. 

9. We have developed very recently (Lou \& Fan 2002; Lou 2002) models 
for stationary fast and slow MHD density waves with logarithmic spirals 
in a magnetized singular isothermal disk (MSID) with a flat rotation 
curve. The Poisson equation can be solved exactly without invoking
the tight-winding approximation (Kalnajs 1971; Shu et al. 2000) so 
that the pitch angle of a logarithmic spiral can be arbitrary. Our 
main motivation is to address the issue of sustaining an extended 
slow MHD density wave pattern within a disk of a flat rotation curve
(Sofue 1996; Ferguson et al. 1998; Frick et al. 2000). Given what 
we have learned from this analysis, it is possible to further
include a CRG into the formalism of Lou \& Fan (2002). 
It is expected that minute enhancements of logarithmic spiral arms 
in the CRG would be phase-shifted relative to enhancements 
of logarithmic spiral arms of magnetic field for either fast and slow 
MHD density waves. An additional stationary MSID configuration with 
logarithmic arms is also expected, corresponding to the suprathermal 
mode (Parker 1965).

\hbox{ }
\vskip 0.4cm
\ctrline{5. NOTES AND DISCUSSIONS }
\vskip 0.4cm

We have studied fast and slow MHD density waves in a rotating magnetized 
thermal gas disk in the presence of CRG that is treated as a tenuous, 
relativistically hot gas fluid for large-scale and low-frequency 
dynamic phenomena. In this formalism, thermal gas and CRG are tied to the 
galactic magnetic field in transverse bulk motions but are allowed to move 
relative to each other along magnetic field lines. One way to
better appreciate fast and slow MHD density waves (Fan \& Lou 1996; Lou 
\& Fan 1998a; Lou et al. 2001a; Lou 2002) is to first understand the basic 
properties of fast and slow MHD waves in a uniformly magnetized medium. 
Similarly, the way to fully appreciate fast and slow MHD density waves in 
the presence of CRG in the current context is to acquaint oneself with the 
basic properties of MHD wave modes in a uniformly magnetized thermal medium 
submerged in a uniform CRG pioneered by Parker (Parker 1965, 1967, 1969; 
see also Appendix B for more specifics). 

The analysis of Parker (1965) leads to following basic theoretical 
facts. First, the conventional slow MHD wave in the thermal gas is 
not very much affected by the presence of CRG, and the 
associated CRG mass density fluctuation, which anti-correlates 
with the parallel magnetic field perturbation (Appendix B), is tiny 
compared to the thermal gas density fluctuation. Secondly, the 
conventional fast MHD wave in the thermal gas, propagating sufficiently 
away from the direction perpendicular to the background magnetic field,
is not very much affected by the presence of CRG, and the 
associated CRG mass density fluctuations, which also anti-correlates 
with the parallel magnetic field perturbation, is also tiny compared 
to the thermal gas density fluctuation. In the propagation direction 
perpendicular to the magnetic field, the conventional fast MHD wave 
is {\it suppressed} (see Fig. 1 of Parker 1965). And finally, there 
exists a suprathermal mode resulting primarily from the compression 
of a relativistically hot CRG with a characteristic sound 
speed close to the speed of light $c$. Enhancements of CRG 
density, parallel magnetic field, and thermal gas density are all in 
phase for the suprathermal mode.

The specific aims of performing this analysis are to derive the
phase relationships between the perturbation enhancements of 
parallel magnetic field and CRG mass density associated with 
fast and slow MHD density waves, to confirm that the phase 
relationships between the perturbation enhancements of parallel 
magnetic field and thermal gas density associated with fast and 
slow MHD density waves in the presence of CRG remain more or less 
the same as in the case where CRG is absent (Fan \& Lou 1996; Lou 
\& Fan 1998a), and to assess the net effect on synchrotron 
radio-continuum emissions as a result of fluctuations in parallel 
magnetic field strength and in CRG mass density.

The basic conclusions of our analysis are the following.
(1) In the presence of CRG, fast and slow MHD 
density waves are only slightly modified, and there exists 
an additional generalized suprathermal mode. 
(2) As expected, the perturbation enhancements of parallel magnetic 
field and the thermal gas are in phase for fast MHD density waves 
and are significantly phase shifted (i.e. $\gsim\pi/2$) for slow MHD 
density waves. 
(3) The perturbation enhancements of the thermal gas and CRG mass 
density are in phase for slow MHD density waves and are signifcantly 
phase shifted (i.e., $\gsim\pi/2$) for fast MHD density waves; in 
other words, the perturbation enhancements of parallel magnetic 
field and CRG mass density are significantly phase shifted (i.e., 
$\gsim\pi/2$) for both fast and slow MHD density waves. 
(4) As CRG mass density fluctuation associated fast and slow MHD 
density waves is tiny, the radio-continuum structures seen in 
synchrotron emissions are primarily determined by magnetic field 
enhancements.

One restriction of our analysis is to require that velocity and magnetic 
field perturbations lie within the gas disk plane. This constraint, which 
leads to a great deal of simplification in the analysis, does exclude
the possible occurrence of Parker instability (i.e., magnetic 
Rayleigh-Taylor instability) in the vertical direction across the 
magnetized disk (Parker 1966, 1967; Shu 1974; Mouschovias 1996 and 
extensive references therein). The Parker instability gives rise to 
magnetic loops or bubbles inflated by cosmic rays on both sides of 
the disk and causes condensation or conglomeration of thermal gas 
clouds into magnetic valleys on spatial scales in the order of 
$\sim 1$ kpc (give-and-take a factor of 2) along the mean magnetic 
field. As a result of vertical undulation of magnetic field lines and 
redistribution of CRG, synchrotron intensity can vary on 
the scale of $\oforder 1\hbox{ kpc}$. Since fast and slow MHD density 
waves vary on spatial scales $\gsim$ several kpc, the overall 
enhancement of synchrotron emission resulting from transverse 
compression of large-scale magnetic field within the disk will not 
be fundamentally altered by relatively small scale magnetic bubbles 
pertruding out of the disk plane. In short, even though quantitative 
calculations of synchrotron contrast must take consequences of 
Parker instability into account, the enhancement of large-scale 
galactic magnetic field plays a dominant role for increasing 
synchrotron radio-continuum intensity.

Another restriction of our analysis is to require that waves
propagate within the disk plane, with vertical wavenumber
$k_z=0$. As noted recently in Lou et al. (2001a), this assumption
excludes a class of magnetorotational instabilities (MRIs) 
given the specified disk and field geometries (e.g., Balbus 
\& Hawley 1992; Terquem \& Papaloizou 1996; Kim \& Ostriker 2000).
The well-known axisymmetric MRIs (Velikhov 1959; Chandrasekhar 
1960, 1961; Balbus \& Hawley 1991, 1998) involve a disk 
differential rotation, a weak magnetic field component parallel 
to the rotation axis, and vertical perturbation variations with
$k_z\neq 0$. For nonaxisymmetric MRIs in a disk with a toroidal 
magnetic field, it is essential to realize the fact $k_z\neq 0$ 
that couples the perturbation equations for velocity and magnetic 
field perturbations {\it perpendicular} to the disk and the 
equations for MHD perturbations {\it coplanar} with the disk 
(Lou et al. 2001a).

\vskip 1.0cm
\noindent
{\bf ACKNOWLEDGMENTS}
\vskip 0.4cm
\noindent
We are grateful to E. N. Parker for insightful discussions. 
This research has been supported in part by grants from
the US NSF (ATM-9320357, AST-9731623) to the University of 
Chicago, by the ASCI Center for Astrophysical Thermonuclear
Flashes at the University of Chicago under Department of
Energy contract B341495, 
by the Visiting Scientist Programs at the Institute 
of Astronomy and Astrophysics, Academia Sinica
(NSC-88-2816-M-001-0010-6 and NSC89-2816-M001-0006-6) and
at the National Taiwan University (NSC89-2112-M002-037),
by the Yangtze Endowment through the Tsinghua University,
by the Special Funds for Major State Basic Science 
Research Projects of China, and by the Collaborative 
Research Fund from the NSF
of China (NSFC) for Young Outstanding Overseas Chinese 
Scholars (NSFC 10028306) at the National Astronomical 
Observatory, Chinese Academy of Sciences.

\hbox{ }
\vskip 1.0cm
\ctrline{REFERENCES}
\vskip 0.4cm



\ref Bardeen J. M., 1975, in Dynamics of Stellar Systems, 
IAU Symposium No. 69, ed. Hayli A., p. 297, Reidel, Dordrecht.

\ref Balbus S. A., Hawley J. F., 1991, ApJ, 376, 214

\ref Balbus S. A., Hawley J. F., 1992, ApJ, 400, 610
 
\ref Balbus S. A., Hawley J. F., 1998, Rev. Mod. Phys., 70, 1








\ref Beck R., Hoernes P., 1996, Nature, 379, 47

\ref Beck R., Brandenburg A., Moss D., Shukurov A., Sokoloff D., 
1996, ARA\&A, 34, 155

\ref Beck, R. et al.
1999, Nature, 397, 324


\ref Berkhuijsen E. M., Wielebinski R., eds, 1978, Structure and 
Properties of Nearby Galaxies, IAU Symposium No. 77, Reidel, Dordrecht

\ref Berkhuijsen E. M., Horellou C., Krause M., Neininger N.,
Poezd A. D., Shukurov A., Sokoloff D. D., 1997, A\&A, 318, 700
\ref Berkhuijsen E. M., Beck R., Walterbos R. A. M., eds. 2000,
232 WE-Heraeus Seminar on The Interstellar Medium in M31 and
M33 (Aachen: Shaker Verlag)

\ref Bertin G., Lin C. C., Lowe S. A., Thurstans R. P., 1989a,
ApJ, 338, 78

\ref Bertin G., Lin C. C., Lowe S. A., Thurstans R. P., 1989b,
ApJ, 338, 104 

\ref Bertin G., Lin, C. C., 1996, Spiral Structure in Galaxies:
A Density Wave Theory. MIT Press, Cambridge, Mass.


\ref Bicay M. D., Helou G., 1990, ApJ, 362, 59
%
\ref Binney J., Tremaine S., 1987, Galactic Dynamics. Princeton 
University Press, Princeton, New Jersey




\ref Cesarsky C. J., 1980, ARA\&A, 18, 289


\ref Chandrasekhar, S. 1960, Proc. Nat.
Acad. Sci., 46, 253

\ref Chandrasekhar S., 1961, Hydrodynamic and Hydromagnetic 
Stability. Dover Publications, New York 





\ref Condon J. J., Anderson M. L., Helou G., 1991, ApJ, 376, 95
\ref Crosthwatte L. P., Turner J. L., 
Ho P. T. P., 2000, AJ, 119, 1720





\ref de Jong T., Klein U., Wielebinski R., 
Wunderlich E., 1985, A\&A, 147, L6
\ref Dickey J. M., Salpeter E. E., 1984, AJ, 284, 461





\ref Elmegreen B. G., 1987, ApJ, 312, 626 


\ref Elmegreen, B. G. 1994, ApJ, 425, L73
%

\ref Elmegreen B. G., 1995, MNRAS, 275, 944

\ref Elmegreen B. G., Elmegreen D. M., Seiden P., 1989, ApJ, 343, 602

\ref Elmegreen D. M., 1981, ApJ Suppl., 47, 229


\ref Elmegreen D. M., Chromey F. R., Warren A. R., 1998, AJ, 116, 2834

\ref Fan Z. H., Lou Y.-Q., 1996, Nature, 383, 800

\ref Fan Z. H., Lou Y.-Q., 1997, MNRAS, 291, 91

\ref Fan Z. H., Lou Y.-Q., 1999, MNRAS, 307, 645

\ref Ferguson A. M. N., Wyse R. F. G., Gallagher J. S., 
Hunter D. A., 1998, ApJ, 506, L19

\ref Frick P., Beck R., Shukurov A., Sokoloff D., 
Ehle M., Kamphuis J., 2000, MNRAS, 318, 925


\ref Galli D., Shu F. H., Laughlin G., Susana L., 2001, ApJ, 551, 367



\ref Goldreich P., Lynden-Bell D., 1965, MNRAS, 130, 97

\ref Goldreich P., Tremaine S., 1978, ApJ, 222, 850
\ref Goldreich P., Tremaine S., 1979, ApJ, 233, 857


\ref Greisen K., 1960, Ann. Rev. Nuclear Science, 10, 62

\ref Han, J. L., Beck R., Ehle M., Haynes R. F., 
Wielebinski R., A\&A, 1999, 348, 405



\ref Harwit M., Pacini F., 1975, ApJ, 200, L127


\ref Helou G., 1991, in The Interpretation of Modern Synthesis Observations
of Spiral Galaxies, ed. N. Duric and Crane P. C.
Astronomical Society of the Pacific Conference Series, vol 18.
pp. 125-133
\ref Helou G., Soifer B. T., Rowan-Robinson M., 1985, ApJ, 298, L7
\ref Helou G., Bicay M. D., 1993, ApJ, 415, 93
\ref Helou G., et al., 1996, A\&A, 315, L157

\ref Hohl F., 1971, ApJ, 168, 343






\ref Jog C. J., Solomon P. M., 1984a, ApJ, 276, 114
\ref Jog C. J., Solomon P. M., 1984b, ApJ, 276, 127
\ref Jog C. J., 1992, ApJ, 390, 378

\ref Jog C. J., 1996, MNRAS, 278, 209


\ref Julian W. H., Toomre A., 1966, ApJ, 146, 810

\ref Kalnajs A. J., 1971, ApJ, 166, 275



\ref Kalnajs A. J., 1973, Proc. Astron. Soc. Australia, 2, 174

\ref Kennicutt R. C. Jr., 1989, ApJ, 344, 685

\ref Kennicutt R. C., Tamblyn P., Congdon C. W., 1994, ApJ, 435, 22

\ref Kent S. M., 1986, AJ, 91, 1301

\ref Kent S. M., 1987, AJ, 93, 816

\ref Kent S. M., 1988, AJ, 96, 514


\ref Kim W.-T., Ostriker E. C., 2000, ApJ, 540, 372

\ref Kormendy J., Norman C. A., 1979, ApJ, 233, 539

\ref Krause M., 1993, in The Cosmic Dynamo, 
IAU Symposium No. 157, eds. Krause F., R\"adler K. H., R\"udiger G. 
pp. 305-310, Kluwer Academic, Dordrecht.

\ref Krause M., Hummel E., Beck R., 1989, A\&A, 217, 4



\ref Lerche I. 1967, ApJ, 149, 395, 533 

\ref Lerche I., Parker E. N., 1966, ApJ, 145, 106

\ref Lin C. C., 1967a, in Relativity Theory and Astrophysics, 
vol. 2, Galactic Structure. ed. Ehlers J., pp. 66-97, Am. Math. 
Soc., Providence, R. I.

\ref Lin C. C., 1967b, ARA\&A, 5, 453

\ref Lin C. C., 1987, Selected Papers of C. C. Lin. 
World Scientific, Singapore.

\ref Lin C. C., Shu F. H., 1964, ApJ, 140, 646

\ref Lin C. C., Shu F. H., 1966, Proc. Natl. Acad. Sci. U.S.A., 55, 229


\ref Lin C. C., Lau Y. Y., 1979, Studies Appl. Math., 60, 97







\ref Lou Y.-Q., 1996a, MNRAS, 279, L67

\ref Lou Y.-Q., 1996b, MNRAS, 279, 129

\ref Lou Y.-Q., 2002, MNRAS, 337, 225

\ref Lou Y.-Q., Fan Z. H., 1997, Communications in 
Nonlinear Science and Numerical Simulation, 2 (No. 2), 59

\ref Lou Y.-Q., Fan Z. H., 1998a, ApJ, 493, 102

\ref Lou Y.-Q., Fan Z. H., 1998b, MNRAS, 297, 84

\ref Lou Y.-Q., Fan Z. H., 2000a, MNRAS, 315, 646

\ref Lou Y.-Q., Fan Z. H., 2000b, in {\it The Interstellar
Medium in M31 and M33}. 232 WE-Heraeus Seminar, eds. E.M.
Berkhuijsen, R. Beck, R.A.M. Walterbos (Aachen: Shaker Verlag),
pp. 205-208

\ref Lou Y.-Q., Fan Z. H., 2002, MNRAS, 329, L62

\ref Lou Y.-Q., Han J. L., Fan Z. H., 1999, MNRAS, 308, L1

\ref Lou Y.-Q., Yuan C., Fan Z. H., 2001, ApJ, 552, 189

\ref Lou Y.-Q., Yuan C., Fan Z. H., Leon S., 2001, ApJ, 553, L35

\ref Lou Y.-Q., Walsh W. M., Han J. L.,
Fan Z. H., 2002, ApJ, 567, 289

\ref Lu N. Y., et al., 1996, A\&A, 315, L153


\ref Lynden-Bell D., 1966, Observatory, 86, 57



\ref Malhotra S., et al., 1996, A\&A, 315, L161



\ref Mathewson D. S., van der Kruit P. C., Brouw W. N., 1972, A\&A, 17, 468

\ref Mestel L., 1963, MNRAS, 126, 553
%




\ref Miller R. H., Prendergast K. H., Quirk W. J., 1970, ApJ, 161, 903

\ref Montenegro L. E., Yuan C., Elmegreen B. G., 1999, ApJ, 520, 592


\ref Mouschovias T. Ch., 1996, in Solar and Astrophysical
Magnetohydrodynamic Flows, NATO ASI series, ed. Tsinganos, K. C. 
pp. 475-504, Kluwer Academic, Dordrecht.


\ref Neininger N., 1992, A\&A, 263, 30

\ref Neininger N., Beck R., Sukumar S., Allen R. J., 1993, A\&A, 274, 687

\ref Ondrechen M. P., 1985, AJ, 90, 1474


\ref Ostriker J. P., Peebles P. J. E., 1973, ApJ, 186, 467


\ref Pacholczyk A. G., 1970, Radio Astrophysics,
Freeman and Company, San Francisco, California
%
%
%
\ref Pannatoni R. F., 1983, Geophys. Astrophys. Fluid Dyn., 24, 165
%
\ref Parker E. N., 1965, ApJ, 142, 1086
\ref Parker E. N., 1966, ApJ, 145, 811
\ref Parker E. N., 1967 in Nebulae and Interstellar Matter
(ed. by L. Aller, D. B. McLaughlin and B. Middlehurst)
University of Chicago Press, Chicago, Vol. VII, Chapter 14
Dynamic Properties of Cosmic Rays. p. 707.
\ref Parker E. N. 1969, Space Sci. Rev., 9, 651
\ref Parker E. N., 1979, Cosmical Magnetic Fields. Clarendon, Oxford
\ref Parker E. N., 1992, ApJ, 401, 137




\ref Quirk W. J., 1972, ApJ, 176, L9





\ref Rix H.-W., Rieke M. J., 1993, ApJ, 418, 123
\ref Roberts W. W., Jr., Yuan C., 1970, ApJ, 161, 887




\ref Safronov V. S., 1960, Ann. d'Astrophysique, 23, 979


%


\ref Scarrott S. M., Ward-Thompson D., Warren-Smith R. F.,
1987, MNRAS, 224, 299



\ref Shu F. H. 1970a, ApJ, 160, 89 

\ref Shu F. H. 1970b, ApJ, 160, 99 
%

\ref Shu, F. H. 1974, A\&A, 33, 55 



\ref Shu F. H., Laughlin G., Lizano S., Galli D., ApJ, 2000, 535, 190

\ref Silk J., 1997, ApJ, 481, 703


\ref Sofue Y., 1996, ApJ, 458, 120





\ref Sukumar S., Allen R. J., 1989, Nature, 340, 537



\ref Terquem C., Papaloizou J. C. B., 1996, MNRAS, 279, 767

\ref Toomre A., 1964, ApJ, 139, 1217


\ref Toomre A., 1977, ARA\&A, 15, 437


\ref Toomre A., Toomre J., 1972, ApJ, 178, 623


\ref Tuffs R. J., et al., 1996, A\&A, 315, L149

\ref Velikhov E. P., 1959, J. Expl. Theoret. Phys.
(U.S.S.R.), 36, 1398


\ref Wang B., Silk J., 1994, ApJ, 427, 759




\ref Woltjer L., 1965, in Galactic Structure, eds. Blaauw A.,
Schmidt M., pp. 531-587, The University of Chicago Press, Chicago
\ref Wunderlich E., Klein U., Wielebinski R., 1987, A\&A, 69, 487






\hbox{ }
\vskip 1.0cm
\ctrline{APPENDIX A}
\vskip 0.4cm

By requiring $u_r=v_r$, the lengthy equation (3.12) for $v_r$ 
contains the information of fast and slow MHD density waves 
and the suprathermal mode (Parker 1965) and can be cast into 
the compact mathematical form of
$$
{\cal C}_1{d^2v_r\over dr^2}+{\cal C}_2{dv_r\over dr}
+{\cal C}_3v_r=0\ ,\eqno(A1)
$$
where the three complex coefficients ${\cal C}_1$,
${\cal C}_2$, and ${\cal C}_3$ are defined by
$$
{\cal C}_1\equiv {iC_A^2\over (\omega-m\Omega)}
+{C_S^2{\cal B}\over
(\mu_{\circ}+\epsilon_{\circ})}
+{i(\omega-m\Omega) C_C^2\Sigma_{\circ}\over
(\mu_{\circ}+\epsilon_{\circ})
[(\omega-m\Omega)^2-m^2C_C^2\Sigma_{\circ}
/(r^2\epsilon_{\circ})]}\ ,\eqno(A2) 
$$
$$\eqalign{
{\cal C}_2\equiv &-{2\Omega\mu_{\circ}
\over (\mu_{\circ}+\epsilon_{\circ})}
\bigg[{r(\omega-m\Omega){\cal B}\over m\mu_{\circ}}-{ir\over m}\bigg]
-{2im\Omega C_C^2\Sigma_{\circ}
\over r(\mu_{\circ}+\epsilon_{\circ})
[(\omega-m\Omega)^2-m^2C_C^2\Sigma_{\circ}
/(r^2\epsilon_{\circ})]}\cr &
+iC_A^2\bigg\lbrace {d\over dr}\bigg({1\over \omega-m\Omega}\bigg)
+r{d\over dr}\bigg[{1\over r(\omega-m\Omega)}\bigg]\bigg\rbrace
-{\cal F}_1{\cal B}+{1\over (\mu_{\circ}+\epsilon_{\circ})}
{d (C_S^2{\cal B})\over dr} \cr &
+ {C_C^2\Sigma_{\circ}
[im\kappa^2/(2\Omega r)+i(\omega-m\Omega)d\ln (r\Sigma_{\circ})
/dr]\over (\mu_{\circ}+\epsilon_{\circ})
[(\omega-m\Omega)^2-m^2C_C^2\Sigma_{\circ}/(r^2\epsilon_{\circ})]}  
+{C_S^2{\cal A}\over (\mu_{\circ}+\epsilon_{\circ})} \cr &
+{1\over (\mu_{\circ}+\epsilon_{\circ})}
{d\over dr}\bigg[{C_C^2\Sigma_{\circ} i(\omega-m\Omega)
\over (\omega-m\Omega)^2-m^2C_C^2\Sigma_{\circ}
/(r^2\epsilon_{\circ})}\bigg] 
-{{\cal B}\over (\mu_{\circ}+\epsilon_{\circ})^2}
{d(p_{\circ}+P_{\circ})\over dr} \cr &
-{\Gamma C_C^2\Sigma_{\circ}/[(\Gamma-1)c^2]\over
(\mu_{\circ}+\epsilon_{\circ})^2}{d(p_{\circ}+P_{\circ})\over dr}
{i(\omega-m\Omega)\over (\omega-m\Omega)^2-m^2C_C^2\Sigma_{\circ}
/(r^2\epsilon_{\circ})}\ ,\cr}\eqno(A3)
$$
$$\eqalign{
{\cal C}_3&\equiv i(\omega-m\Omega)-{2\Omega\mu_{\circ}\over
(\mu_{\circ}+\epsilon_{\circ})}\bigg[{r(\omega-m\Omega )\over
m\mu_{\circ} }{\cal A}
-{i\over m\mu_{\circ}}{d(r\mu_{\circ})\over dr}\bigg]\cr &
-{2\Omega\epsilon_{\circ}\over (\mu_{\circ}+\epsilon_{\circ})}
\bigg [{mC_C^2\Sigma_{\circ}\over r(\omega-m\Omega )\epsilon_{\circ}}
{im\kappa^2/(2\Omega r)+i(\omega-m\Omega)d\ln (r\Sigma_{\circ})
/dr\over (\omega-m\Omega)^2-m^2C_C^2\Sigma_{\circ}/
(r^2\epsilon_{\circ})}\cr &
+{i\kappa^2\over 2\Omega (\omega-m\Omega)}\bigg ]
-{im^2 C_A^2\over r^2(\omega-m\Omega)}
+iC_A^2{d\over dr}\bigg\lbrace r{d\over dr}\bigg[
{1\over r(\omega-m\Omega)}\bigg]\bigg\rbrace \cr &
-{\cal F}_1{\cal A}
+{1\over (\mu_{\circ}+\epsilon_{\circ})}{d(C_S^2{\cal A})\over dr} 
-{{\cal A}\over (\mu_{\circ}+\epsilon_{\circ})^2}
{d(p_{\circ}+P_{\circ})\over dr} \cr &
+{1\over (\mu_{\circ}+\epsilon_{\circ})}
{d\over dr}\bigg\lbrace {C_C^2\Sigma_{\circ}
[im\kappa^2/(2\Omega r)+i(\omega-m\Omega)d\ln (r\Sigma_{\circ})
/dr]\over (\omega-m\Omega)^2-m^2C_C^2\Sigma_{\circ}
/(r^2\epsilon_{\circ})}\bigg\rbrace  \cr &
-{\Gamma C_C^2\Sigma_{\circ}/[(\Gamma -1)c^2]\over
(\mu_{\circ}+\epsilon_{\circ})^2}{d(p_{\circ}+P_{\circ})\over dr}
{[im\kappa^2/(2\Omega r)+i(\omega-m\Omega)d\ln (r\Sigma_{\circ})
/dr]\over (\omega-m\Omega)^2-m^2C_C^2\Sigma_{\circ}/(r^2\epsilon_{\circ})}
\ .\cr}\eqno(A4)
$$
The complex coefficients ${\cal F}_1$, ${\cal A}$, and ${\cal B}$ 
are defined by equations (3.5) and (3.6) in the main text.

\hbox{ }
\vskip 0.4cm
\ctrline{APPENDIX B}
\vskip 0.4cm

For the convenience of reference and discussion, we derive and 
summarize here several pertinent phase relationships 
among perturbation variables, with a generalization of the 
original analysis by Parker (1965) for hydromagnetic perturbations 
in a uniformly magnetized medium of thermal and suprathermal gases. 
To accommodate the possible parameter regime of
$W\equiv B/[4\pi (\rho_{\circ}+\delta_{\circ})]^{1/2}\gsim c$ with
$c$ being the speed of light, we include the effect of 
displacement current perturbation for the Lorentz force term in the 
momentum equation. Otherwise, the Cartesian coordinates $(x,y,z)$, 
notations, and definitions all remain essentially the same as those 
adopted by Parker (1965).

From the mass conservation and $z-$component of the
momentum equation for the thermal gas, one has
$$
u_y={\rho(\omega^2-k^2a^2\cos^2\theta)
\over\omega\rho_{\circ}k\sin\theta}\ ,\eqno(B1)
$$
where $u_y$ is the bulk velocity perturbation of the thermal 
gas perpendicular to the background magnetic field $B\hat z$, 
$\omega$ is the angular frequency, $k$ is the total wavenumber, 
$\theta$ is the angle in the $yz-$plane between the wave vector 
$\vec k$
and the background magnetic field $B\hat z$, $a$ is the sound 
speed of the thermal gas, $\rho_{\circ}$ is the uniform 
background thermal gas density, and $\rho$ is the thermal gas 
density perturbation. In parallel, from the mass conservation 
and $z-$component of the momentum equation for the suprathermal 
gas, one has
$$
v_y={\delta(\omega^2-k^2b^2\cos^2\theta)
\over\omega\delta_{\circ}k\sin\theta}\ ,\eqno(B2)
$$
where $v_y$ is the bulk velocity perturbation of the suprathermal 
gas perpendicular to the background magnetic field $B\hat z$,
$b$ is the sound speed in the suprathermal gas, $\delta_{\circ}$ 
is the uniform background suprathermal gas density, and $\delta$ 
is the suprathermal gas density perturbation. By the requirement 
that the bulk velocities of thermal gas and superathermal gas are 
the {\it same} transverse to the background magnetic field $B\hat z$, 
namely $u_y=v_y$, we derive from equations $(B1)$ and $(B2)$ 
$$
{\delta\over\delta_{\circ}}=
{\rho(\omega^2-k^2a^2\cos^2\theta)\over\rho_{\circ}
(\omega^2-k^2b^2\cos^2\theta)}\ ,  \eqno(B3)
$$
where $\delta/\delta_{\circ}$ and $\rho/\rho_{\circ}$ are the 
relative mass density fluctuations in the suprathermal and 
thermal gases, respectively.

By combining the $z-$component of the magnetic induction equation
$$
\omega b_z=Bk\sin\theta u_y\ \eqno(B4)
$$
with equation $(B1)$ where $b_z$ is the $z-$component of the
magnetic field perturbation, one has the relative fluctuation in 
parallel magnetic field $b_z/B$ and the relative mass density 
fluctuation in thermal gas $\rho/\rho_{\circ}$ related by
$$
{b_z\over B}=
{\rho(\omega^2-k^2a^2\cos^2\theta)\over\rho_{\circ}\omega^2}\ .\eqno(B5)
$$
The incompressible transverse Alfv\'en mode involves $u_x=v_x$ and 
$b_x\neq 0$, and solution (9) of Parker (1965) for its dispersion 
relation now takes the form of
$$
\omega^2={k^2W^2\cos^2\theta\over (1+W^2/c^2)}\ ,\eqno(B6)
$$
where $W\equiv B/[4\pi(\rho_{\circ}+\delta_{\circ})]^{1/2}$.
The modified dimensionless dispersion relation for 
the fast, slow, and suprathermal modes now becomes 
$$\eqalign{
(U^2-M^2)&(U^2-\cos^2\theta)(U^2-n^2\cos^2\theta)\cr &
-{U^2\sin^2\theta
[(1+\alpha)U^2-(1+\alpha/n^2)n^2\cos^2\theta]
\over (1+\alpha/n^2)(1+W^2/c^2)}=0\ ,\cr}\eqno(B7)
$$
where $U\equiv \omega/(ka)$, $M^2\equiv W^2/[a^2(1+W^2/c^2)]$, 
$n\equiv b/a$, 
$\alpha\equiv\delta_{\circ}b^2/(\rho_{\circ} a^2)$ (for a comparison,
see Parker's eqn [15] and the relevant definitions). The factor 
$(1+W^2/c^2)$ that appears in various places is due to the 
displacement current effect. We now examine systematically the phase 
relationships among various perturbation enhancements such as 
$\rho/\rho_{\circ}$, $b_z/B$, and $\delta/\delta_{\circ}$ for the 
fast, slow, and suprathermal modes separately, using the phase 
relations $(B3)$ and $(B5)$.

\noindent
1. The special case of $\theta=0$ (parallel wave
propagations relative to $\vec B$). 

Dispersion relation $(B7)$ gives three exact solutions, 
$$
U^2=M^2\ ,\eqno(B8)
$$
$$
U^2=1\ ,\eqno(B9)
$$
$$
U^2=n^2\ .\eqno(B10)
$$
For the modified Alfv\'en wave speed $C_D\equiv W/(1+W^2/c^2)^{1/2}$ 
greater (or less) than the thermal sound speed $a$, solution $(B8)$ 
corresponds to the fast (or slow) mode, while solution $(B9)$ 
corresponds to the slow (or fast) mode. Solution $(B10)$ represents 
the suprathermal mode.

For the suprathermal mode $(B10)$ with $\omega^2=k^2b^2$, one gets 
$\rho=0$, $p=0$, $u_z=0$, $u_y=0$, $\delta\neq 0$, $P\neq 0$, 
$v_z\neq0$, $b_y=0$, and $b_z=0$. Physically, this is an acoustic 
wave in the suprathermal gas propagating along the background 
magnetic field $\vec B$ without disturbing the thermal gas. For a 
relativistically hot CRG, the value of $b$ can be fairly close to 
the speed of light $c$.

For $C_D>a$, the slow mode $(B9)$ with $\omega^2=k^2a^2$ gives
$\rho\neq 0$, $p\neq 0$, $u_z\neq 0$, $u_y=0$, $\delta=0$, $P=0$, 
$v_z=0$, $b_y=0$, and $b_z=0$. Physically, this is an acoustic 
wave in the thermal gas propagating along the background magnetic 
field $\vec B$ without disturbing the suprathermal gas.

For $C_D>a$, the fast mode $(B8)$ with $\omega^2=k^2W^2$ involves 
all nonzero perturbation variables. By equation $(B3)$, 
$\rho/\rho_{\circ}$ and $\delta/\delta_{\circ}$ are out of phase, 
while by equation $(B5)$, $\rho/\rho_{\circ}$ and $b_z/B$ are in 
phase as expected.

For $C_D<a$, the fast mode $(B9)$ with $\omega^2=k^2a^2$ has the 
same characteristics of an acoustic wave in the thermal gas 
propagating along the background magnetic field. One simply 
changes the name of this mode because of a faster wave speed.

For $C_D<a$, the slow mode $(B8)$ with $\omega^2=k^2W^2$ involves
all nonzero perturbation variables. By equation $(B3)$,
$\rho/\rho_{\circ}$ and $\delta/\delta_{\circ}$ are now in phase,
while by equation $(B5)$, $\rho/\rho_{\circ}$ and $b_z/B$ are now 
out of phase as expected.

\noindent
2. The case of $\theta\sim\pi/2$ (nearly perpendicular 
wave propagations relative to $\vec B$).

Here, we would first correct typos in Parker's paper in the 
case of $\theta$ being sufficiently close to $\pi/2$: Parker's 
solution (19)
$$
U^2={M^2\over M^2+1}\cos^2\theta \eqno(B11)
$$
should be associated with the {\it minus} (not plus)
sign of $R$ for the slow mode, while his solution (20)
$$
U^2\cong n^2\cos^2\theta {M^2+1\over M^2+1+\alpha} \eqno(B12)
$$
should be associated with the {\it plus} (not minus) sign 
of $R$ for the fast mode, where $R$ is explicitly defined 
by his equation (18), namely
$$
R\equiv \pm [(M^2+1)^2-4M^2\cos^2\theta]^{1/2}\ \eqno(B13)
$$ 
(see his eqn. [15] or our eqn. [$B7$] with $W^2/c^2\ll 1$).
Also, a more accurate version of his solution (21) for the 
suprathermal mode close to $\theta=\pi/2$ would be
$$
U^2\cong M^2+{\alpha+1\over 1+\alpha/n^2}+
n^2\cos^2\theta {\alpha\over M^2+\alpha+1}\ ,\eqno(B14)
$$
as can be readily seen from his equation $(15)$ by setting 
$\theta=\pi/2$.

In the presence of displacement current effects, one can also
derive three approximate solutions from dispersion relation 
$(B7)$ that may be written in the polynomial form of
$$\eqalign{
&U^6-\bigg[(n^2+1)\cos^2\theta+M^2
+{(1+\alpha)\sin^2\theta
\over (1+\alpha/n^2)(1+W^2/c^2)}\bigg]U^4\cr &
+\cos^2\theta\bigg[\bigg(\cos^2\theta+{\sin^2\theta
\over 1+W^2/c^2}\bigg)n^2+M^2(n^2+1)\bigg]U^2
-M^2n^2\cos^4\theta=0\ ,\cr}\eqno(B15)
$$
where $M^2\equiv W^2/[a^2(1+W^2/c^2)]$. For $\theta\sim\pi/2$ and 
$n^2\gg 1$, the dispersion relation of the slow mode now becomes
$$
U^2\cong {M^2\cos^2\theta\over M^2(1+1/n^2)
+\sin^2\theta/(1+W^2/c^2)+\cos^2\theta }\ ,\eqno(B16)
$$
the dispersion relation of the fast mode now becomes
$$
U^2\cong {\cos^2\theta\{[\cos^2\theta
+\sin^2\theta/(1+W^2/c^2)]n^2+M^2(n^2+1)\}
\over (n^2+1)\cos^2\theta+M^2+
(1+\alpha)\sin^2\theta/[(1+\alpha/n^2)(1+W^2/c^2)]}\ ,
\eqno(B17)
$$
and the dispersion relation of the 
suprathermal mode now becomes
$$
U^2\cong M^2+{\alpha+1\over (1+\alpha/n^2)(1+W^2/c^2)}+
{\alpha n^2\cos^2\theta\over M^2+\alpha+1}\ .\eqno(B18)
$$

For the slow mode $(B16)$, $b_z/B$ and $\rho/\rho_{\circ}$
are out of phase by using equation $(B5)$ as expected, while 
$\delta/\delta_{\circ}$ and $\rho/\rho_{\circ}$ are in phase 
by using equation $(B3)$.

For the fast mode $(B17)$, $b_z/B$ and $\rho/\rho_{\circ}$
are in phase by using equation $(B5)$, while 
$\delta/\delta_{\circ}$ and $\rho/\rho_{\circ}$ are out of 
phase by using equation $(B3)$.

For the suprathermal mode $(B18)$, $b_z/B$ and $\rho/\rho_{\circ}$ 
are in phase by using equation $(B5)$, and $\delta/\delta_{\circ}$ 
and $\rho/\rho_{\circ}$ are also in phase by using equation $(B3)$.

\noindent
3. The case of $n^2\cos^2\theta$ being 
sufficiently larger than 1 with $\theta\neq 0$.

Corresponding to the suprathermal mode of Parker's solution (17),
we now have
$$
U^2=n^2\cos^2\theta+{\alpha\sin^2\theta\over 1+W^2/c^2}
+{\cal O}\bigg({1\over n^2\cos^2\theta}\bigg)\ ,\eqno(B19)
$$
and one can readily show that
$\delta/\delta_{\circ}$ and $\rho/\rho_{\circ}$ are in phase 
by equation $(B3)$, and $b_z/B$ and $\rho/\rho_{\circ}$ are 
also in phase by equation $(B5)$.

Corresponding to the fast and slow modes 
of Parker's solution (16), we now have 
$$\eqalign{
U^2=&{[M^2+\cos^2\theta+\sin^2\theta/(1+W^2/c^2)+R_D]\over 2}\cr &
\times\bigg\lbrace 1-{\alpha [M^2+\cos^2\theta
+\sin^2\theta/(1+W^2/c^2)+R_D]\tan^2\theta\over 
2R_Dn^2(1+W^2/c^2)}\bigg\rbrace \cr &\qquad\qquad\qquad
+{\cal O}\bigg({1\over n^4\cos^4\theta}\bigg)\ ,\cr}\eqno(B20)
$$
where, replacing definition $(B13)$ for $R$, $R_D$ is defined by 
$$
R_D\equiv \pm \{[M^2+\cos^2\theta+\sin^2\theta/(1+W^2/c^2)]^2
-4M^2\cos^2\theta\}^{1/2}\ \eqno(B21)
$$
with the plus and minus signs of $R_D$ corresponding 
to the fast and slow modes, respectively.

For the fast mode $(B20)$ with the plus sign of $R_D$ in 
definition $(B21)$, one can show that $\delta/\delta_{\circ}$ 
and $\rho/\rho_{\circ}$ are out of phase by equation $(B3)$,
while $b_z/B$ and $\rho/\rho_{\circ}$ are in phase by
equation $(B5)$ as expected.

Meanwhile, for the slow mode $(B20)$ with the minus sign of 
$R_D$ in definition $(B21)$, one can show that $\delta/\delta_{\circ}$ 
and $\rho/\rho_{\circ}$ are in phase by equation $(B3)$, while 
$b_z/B$ and $\rho/\rho_{\circ}$ are out of phase by equation 
$(B5)$ as expected.

We now summarize the results for the parameter regime of 
$n\gg 1$ (i.e., $b\gg a$) with comparable $W$ and $a$. 

\noindent
1. For the suprathermal mode, $\delta/\delta_{\circ}$, 
$\rho/\rho_{\circ}$, and $b_z/B$ are all in phase.

\noindent
2. For the fast mode, $\delta/\delta_{\circ}$ and $b_z/B$ are 
out of phase with $\rho/\rho_{\circ}$ and $b_z/B$ being in phase.

\noindent
3. For the slow mode, $\delta/\delta_{\circ}$ and $b_z/B$ are out 
of phase with $\rho/\rho_{\circ}$ and $b_z/B$ being out of phase.

\hbox{ }
\vskip 0.4cm
\ctrline{APPENDIX C}
\vskip 0.4cm     
%

The mass conservation for the interstellar medium (ISM) is
$$
{\partial\rho\over\partial t}+\nabla\cdot (\rho\vec v)=0\ ,
\eqno(C1)
$$
where $\rho$ is the ISM mass density and $\vec v$ is the 
bulk flow velocity of the ISM. After the operation of 
vertical integration $\int dz$, one has
$$
{\partial\mu\over\partial t}+\nabla\cdot (\mu\vec v)=0\ ,
\eqno(C2)
$$    
where $\mu\equiv \int dz\rho$ is the surface mass density
of the ISM.

Similarly, the mass conservation for the cosmic-ray gas 
(CRG) is
$$
{\partial\delta\over\partial t}+\nabla\cdot (\delta\vec u)=0\ ,
\eqno(C3)
$$
where $\delta$ is the mass density and $\vec u$ is the bulk
flow velocity of the CRG. After the operation of vertical
integration $\int dz$, one has
$$
{\partial\Sigma\over\partial t}+\nabla\cdot (\Sigma\vec u)=0\ ,
\eqno(C4)
$$
where $\Sigma\equiv \int dz\delta$ is the surface mass density
of the CRG.    

We invoke the polytropic approximations for both the ISM
and the CRG, 
$$
p=C_S^2\mu\  \eqno(C5)
$$
and
$$
P=C_C^2\Sigma\ ,\eqno(C6)
$$
where $C_S$ is the thermal sound speed of the ISM and $C_C$ is 
the effective sound speed of the CRG, $p$ is the vertically 
integrated (two-dimensional) gas pressure of the ISM, and $P$ is 
the vertically integrated (two-dimensional) effective pressure 
of the CRG which is regarded as a ``relativistically hot" gas.

For a plasma of infinite conductivity, 
the magnetic induction equation is 
$$
{\partial\vec B\over\partial t}=\nabla\times(\vec v\times\vec B)\ ,
\eqno(C7)
$$
where the magnetic field $\vec B$ satisfies 
the divergence-free condition of 
$$
\nabla\cdot\vec B=0\ .\eqno(C8)
$$
In the current context, bulk flow velocities $\vec v_{\perp}$
and $\vec u_{\perp}$ perpendicular to $\vec B$ are constrained 
to be the same by the large-scale mean magnetic field.

By ignoring effects of displacement current 
for nonrelativistic flows, we have
$$
\nabla\times\vec B={4\pi\over c}(\vec J_{ISM}+\vec J_{CRG})\ ,
\eqno(C9)
$$
where $c$ is the speed of light, $\vec J_{ISM}$ and $\vec J_{CRG}$ 
are the electric current densities present in the ISM and in the 
CRG, respectively.

The (two-dimensional) momentum equation for the ISM is
$$
\mu {D\vec v\over Dt}+\nabla p
-\int dz {\vec J_{ISM}\times\vec B\over c}
-\mu\nabla\Phi=0\ ,\eqno(C10)
$$
and the (two-dimensional) momentum equation for the CRG is
$$
\Sigma {D\over Dt}\bigg({\psi\vec u\over \bar m c^2}\bigg)
+\nabla P-\int dz{\vec J_{CRG}\times\vec B\over c}
-{\Sigma\psi\over\bar mc^2}\nabla\Phi=0\ ,\eqno(C11)
$$  
where $\bar m$ is the mean particle mass in the CRG, $\Phi$ is 
the negative gravitational potential, $\psi\equiv \bar mc^2+
\Gamma\bar m P/[(\Gamma-1)\Sigma]$ is the proper specific 
enthalpy including that of the particle rest mass. Given the 
polytropic approximation, $\psi$ is a constant.

For the background rotational equilibrium in the radial direction, 
the above two equations can be combined to derive equation (2.1) 
(denoting $\epsilon\equiv\Sigma\psi/[\bar m c^2]$). Likewise, 
perturbations of these two equations can be combined to derive 
the radial component equation (2.6) by using equation (2.1) and 
the force-free approximation. 

\par\vfill\end